\documentclass[journal]{IEEEtran}
\IEEEoverridecommandlockouts
\usepackage[T1]{fontenc}
\usepackage{url}
\usepackage{cite}
\usepackage{graphicx}
\usepackage{amssymb}
\usepackage{eucal}
\usepackage{color}
\usepackage{comment}
\usepackage{arydshln}
\usepackage{authblk}
\usepackage{algpseudocode}
\usepackage{xcolor}
\algnewcommand\algorithmicinput{\textbf{Input:}}
\algnewcommand\INPUT{\item[\algorithmicinput]}
\algnewcommand\algorithmicoutput{\textbf{Output:}}
\algnewcommand\OUTPUT{\item[\algorithmicoutput]}
\usepackage{mathtools}

\usepackage{enumerate}
\usepackage[ruled,vlined]{algorithm2e}
\usepackage{float}

\usepackage{amsthm}
\theoremstyle{definition}
\newtheorem{definition}{Definition}

\newtheorem{lemma}{Lemma}
\newtheorem{theorem}{Theorem}
\newtheorem{corollary}{Corollary}

\newtheorem{example}{Example}
\newtheorem{construction}{Construction}
\newtheorem{observation}{Observation}

\renewcommand{\qed}{\hfill$\blacksquare$}

\newcommand{\cC}{\mathcal{C}}
\newcommand{\cR}{\mathcal{R}}

\newcommand{\remove}[1]{}
\usepackage{tikz}
\usetikzlibrary{positioning}

\newcommand\nc\newcommand
\nc{\vzero}{{\boldsymbol{0}}}
\nc{\bfc}{{\boldsymbol c}}
\nc\bfr{{\boldsymbol r}}
\nc\bfa{{\boldsymbol a}}\nc\bfA{{\boldsymbol A}}\nc\cA{{\mathcal A}}
\nc\bfb{{\boldsymbol b}}\nc\bfB{{\boldsymbol B}}\nc\cB{{\mathcal B}}
\nc\bfd{{\boldsymbol d}}\nc\bfD{{\boldsymbol D}}\nc\cD{{\mathcal D}}
\nc\bfe{{\boldsymbol e}}\nc\bfE{{\boldsymbol E}}\nc\cE{{\mathcal E}}
\nc\bff{{\boldsymbol f}}\nc\bfF{{\boldsymbol F}}\nc\cF{{\mathcal F}}
\nc\bfg{{\boldsymbol g}}\nc\bfG{{\boldsymbol G}}\nc\cG{{\mathcal G}}
\nc\bfh{{\boldsymbol h}}\nc\bfH{{\boldsymbol H}}\nc\cH{{\mathcal H}}
\nc\bfi{{\boldsymbol i}}\nc\bfI{{\boldsymbol I}}\nc\cI{{\mathcal I}}
\nc\bfj{{\boldsymbol j}}\nc\bfJ{{\boldsymbol J}}\nc\cJ{{\mathcal J}}
\nc\bfk{{\boldsymbol k}}\nc\bfK{{\boldsymbol K}}\nc\cK{{\mathcal K}}
\nc\bfl{{\boldsymbol l}}\nc\bfL{{\boldsymbol L}}\nc\cL{{\mathcal L}}
\nc\bfm{{\boldsymbol m}}\nc\bfM{{\boldsymbol M}}\nc\cM{{\mathcal M}}
\nc\bfn{{\boldsymbol n}}\nc\bfN{{\boldsymbol N}}\nc\cN{{\mathcal N}}
\nc\bfo{{\boldsymbol o}}\nc\bfO{{\boldsymbol O}}\nc\cO{{\mathcal O}}
\nc\bfp{{\boldsymbol p}}\nc\bfP{{\boldsymbol P}}\nc\cP{{\mathcal P}}
\nc\bfq{{\boldsymbol q}}\nc\bfQ{{\boldsymbol Q}}\nc\cQ{{\mathcal Q}}
\nc\bfs{{\boldsymbol s}}\nc\bfS{{\boldsymbol S}}\nc\cS{{\mathcal S}}
\nc\bft{{\boldsymbol t}}\nc\bfT{{\boldsymbol T}}\nc\cT{{\mathcal T}}
\nc\bfu{{\boldsymbol u}}\nc\bfU{{\boldsymbol U}}\nc\cU{{\mathcal U}}
\nc\bfv{{\boldsymbol v}}\nc\bfV{{\boldsymbol V}}\nc\cV{{\mathcal V}}
\nc\bfw{{\boldsymbol w}}\nc\bfW{{\boldsymbol W}}\nc\cW{{\mathcal W}}
\nc\bfx{{\boldsymbol x}}\nc\bfX{{\boldsymbol X}}\nc\cX{{\mathcal X}}
\nc\bfy{{\boldsymbol y}}\nc\bfY{{\boldsymbol Y}}\nc\cY{{\mathcal Y}}
\nc\bfz{{\boldsymbol z}}\nc\bfZ{{\boldsymbol Z}}\nc\cZ{{\mathcal Z}}

\newcommand{\ZZ}{\mathbb{Z}}

\nc{\dist}{{\rm d}}

\nc{\syntime}{\mathsf{time}}
\nc{\cycle}{\mathsf{cycle}}
\nc{\strand}{\mathsf{strand}}
\nc{\sC}{\mathsf{C}}
\nc{\syndef}{\mathsf{SynDef}}
\nc{\del}{\mathsf{Del}}
\nc{\ball}{\mathsf{Ball}}
\nc{\ballsyndel}{\ball^\syndel}
\nc{\balldel}{\ball^\del}
\nc{\VT}{{\rm VT}}
\nc{\SVT}{{\rm SVT}}
\nc{\dec}{{\rm DEC}}
\nc{\etal}{{\em et al.}}
\nc{\diff}{{\rm Diff}}
\nc{\rll}{{\rm RLL}}
\nc{\Sum}{{\rm Sum}}

\usepackage{hyperref}
\newcommand\myshade{70}
\hypersetup{
	linkcolor  = red!\myshade!black,
	citecolor  = blue!\myshade!black,
	urlcolor   = blue!\myshade!black,
	colorlinks = true,
}

\title{Coding for Synthesis Defects}
\author{Ziyang~Lu, Han~Mao~Kiah, Yiwei~Zhang, Robert~N.~Grass, and Eitan~Yaakobi
\thanks{Ziyang~Lu and Yiwei~Zhang are with Key Laboratory of Cryptologic Technology and Information Security of Ministry of Education, School of Cyber Science and Technology, Shandong University, Qingdao, Shandong, 266237, China (e-mail: \href{mailto:zylu@mail.sdu.edu.cn}{zylu@mail.sdu.edu.cn}, \href{mailto:ywzhang@sdu.edu.cn}{ywzhang@sdu.edu.cn}).}
\thanks{Han~Mao~Kiah is with the School of Physical and Mathematical Sciences, Nanyang Technological University, Singapore (e-mail: \href{mailto:hmkiah@ntu.edu.sg}{hmkiah@ntu.edu.sg})}
\thanks{Robert~N.~Grass is with the Department of Chemistry and Applied Biosciences, ETH Z\"urich, Vladimir-Prelog-Weg 1-5, 8093 Z\"urich, Switzerland (e-mail: \href{mailto:robert.grass@chem.ethz.ch}{robert.grass@chem.ethz.ch})}
\thanks{Eitan~Yaakobi is with the Department of Computer Science, Technion - Israel Institute of Technology, Haifa 3200003, Israel (e-mail: \href{mailto:yaakobi@cs.technion.ac.il}{yaakobi@cs.technion.ac.il})}
}

\date{}

\begin{document}

\maketitle

\begin{abstract}
Motivated by DNA based data storage system, we investigate the errors that occur when synthesizing DNA strands in parallel, where each strand is appended one nucleotide at a time by the machine according to a template supersequence. If there is a cycle such that the machine fails, then the strands meant to be appended at this cycle will not be appended, and we refer to this as a \emph{synthesis defect}. In this paper, we present two families of codes correcting synthesis defects, which are \emph{$t$-known-synthesis-defect correcting codes} and \emph{$t$-synthesis-defect correcting codes}. For the first one, it is assumed that the defective cycles are known, and each of the codeword is a quaternary sequence. We provide constructions for this family of codes for $t=1,2$, with redundancy $\log4$ and $\log n+18\log3$, respectively. For the second one, the codeword is a set of $M$ ordered sequences, and we give constructions for $t=1,2$ to show a strategy for constructing this family of codes. Finally, we derive a lower bound on the redundancy for single-known-synthesis-defect correcting codes, which assures that our construction is almost optimal.
\end{abstract}

\section{Introduction}
    Storing digital information on synthetic DNA strands has attracted significant interest due to its potential for information storage, particularly its durability and high storage density (see~\cite{yazdi2015dna,shomorony2022information} and references therein).
    The storage process involves converting binary digital information into quaternary strings (nucleotide bases) and writing these onto DNA strands using a synthesis machine.

    While numerous experiments demonstrated the feasibility of DNA storage (see Table 1.1 in \cite{shomorony2022information} for a recent survey), practical implementation for large-scale data remains challenging primarily due to its high cost. Specifically, DNA synthesis stands out as the most costly component in the storage model (see \cite{yu2024high} for a discussion on synthesis procedures).
    Therefore, understanding the synthesis process is essential to enhance efficiency and reduce costs.

    To minimize errors during the synthesis process, DNA strands typically contain no more than 250 nucleotides. Consequently, the user split the encoded quaternary strings into multiple short sequences for synthesis, storing them in an unordered manner. Synthesis of these multiple strands is typically array-based or performed in parallel (see for example,~\cite[Fig.~6]{yu2024high}).
    
    In this process, the synthesis machine scans a fixed template DNA sequence and appends one nucleotide at a time according to this template supersequence. Each synthesis cycle involves appending a nucleotide to a subset of DNA strands requiring that specific nucleotide. Therefore, all synthesized DNA strands must be subsequences of the fixed template supersequence. The length of the fixed supersequence dictates the number of cycles needed for strand generation, thereby influencing the total synthesis time. Minimizing the total synthesis time is essential for reducing the overall cost of DNA synthesis, as each cycle requires reagents and chemicals \cite{ceze2019molecular}.
    This results in an interesting constrained coding problem that was initiated by Lenz \etal~\cite{lenz2020coding}
    and subsequently studied in~\cite{lenz2021multivariate, makarychev2022batch,elishco2023optimal, abu2023dna, chrisnata2023deletion}.

    In this paper, we introduce a new error model that arises during this synthesis process \cite{lietard2021chemical}.
    
    Specifically, if the synthesis machine fails to append a nucleotide at a certain cycle, then we refer to this error event as a {\em synthesis defect} (see Definition~\ref{def:syndef} for a formal definition).
    While synthesis defects share certain similarities to deletions, our primary contribution in this study is to demonstrate that by leveraging certain side information, we can dramatically reduce the redundancy of the coding scheme.
    Our results are distinguished between two scenarios: one where the defective cycles are known, and  another where the defective cycles are unknown. In the first case, we construct codes that correct one and two synthesis defects using $\log 4$ and $\log n + O(1)$ redundant bits, respectively, where $n$ denotes the codeword length.
    In contrast, if we employ a single and a two-deletion-correcting code, we require at least $\log n$ and $2\log n$ redundant bits, respectively. Also, in this case where the defects are known and $t=1$, we provide a matching lower bound for the redundancy.
    For the second case, where the defective cycles are not known, we consider the task of synthesizing $M$ length-$n$  words. Here, we construct codes that correct one and two synthesis defects using roughly $\lambda_1(\log n)^2 +M \log \log n$ and $\lambda_2(\log n)^2 + 2M\log n$ redundant bits, respectively, for some constants $\lambda_1$ and $\lambda_2$.

\section{Problem Formulation}

Throughout this paper, we use $\Sigma=\{1,2,3,4\}$ to denote DNA alphabet of size four and consider the shifted modulo operator so that $(a\bmod 4)$ always belongs to $\Sigma$. For an integer $T$, we use $[T]$ to denote the set $\{1,2,\ldots, T\}$. For a length-$n$ sequence $\bfx$, its $i$th symbol is denoted as $x_i$, and we use $x_1 x_2 \ldots x_n$ or $(x_1,x_2,\ldots,x_n)$ to represent it. Denote $\bfx_{I}=x_{i_1}x_{i_2}\ldots x_{i_k}$ as a subsequence of $\bfx$ indexed by $I$, where $I=\{i_1,i_2,\ldots,i_k\}\subseteq[n]$ is a set of indices of size $k$.

Consider synthesizing a quaternary sequence $\bfx=x_1 x_2 \ldots x_n \in \Sigma^n$. In this paper, we consider the template sequence $12341234\cdots$ and appends one nucleotide according to this sequence in each cycle. Thus, in almost all cases, more than $n$ synthesis cycles are required as we need to wait for the synthesis machine to append the corresponding nucleotide.

Now, to determine the number of cycles, we perform the following computation.
First, we define the \textit{difference sequence} of $\bfx$ to be $\diff(\bfx)=(x_1, x_2-x_1\bmod4,\ldots,x_n-x_{n-1}\bmod4$). Note that the map $\diff:\Sigma^n\rightarrow\Sigma^n$ is a one-to-one mapping and hence it is invertible.

Next, we use $\cycle(\bfx)\in[4n]^n$ to denote the \textit{synthesis cycle sequence} of $\bfx$.
Here, for $1\leq i\leq n$, the $i$th symbol of $\bfx$ will be synthesized in the cycle given by $\cycle(\bfx)_i=\sum_{j=1}^{i}\diff(\bfx)_j$.
For example, if $\bfx=1241321$, then $\diff(\bfx)=1121233$ and $\cycle(\bfx)=(1,2,4,5,7,10,13)$. Observe that the elements of $\cycle(\bfx)$ are necessarily monotonically increasing. Hence, by a slight abuse of notation, we treat $\cycle(\bfx)$ both as a set and as a sequence. Specifically, sometimes, we write $\Delta\cap \cycle(\bfx)$ to mean $\Delta\cap \{\cycle(\bfx)_i\}_{i=1}^n$.

Finally, for a subset of synthesis cycles $\Delta\subseteq[4n]$, we let $I(\bfx, \Delta)=\{i\in [n]:\cycle(\bfx)_i\in\Delta\}$ represent the indices of $\bfx$ whose synthesis cycles belong to $\Delta$. Then, we define $\syndef_\Delta(\bfx)\triangleq\bfx_{[n]\setminus I(\bfx, \Delta)}$ to be the sequence obtained by deleting the symbols whose synthesis cycles belong to $\Delta$.
Let us continue our example with $\bfx=1241321$. When $\Delta=\{12,13\}$, then $I(\bfx, \Delta)=\{7\}$ and $\syndef_\Delta(\bfx)=124132$.

\subsection{Problem Formulation}

In this paper, we consider the scenario where a set of DNA strands is synthesized in parallel.
Similar to before, in every cycle, the machine appends the corresponding nucleotide to the strands that need it.
However, if the machine does not append a nucleotide at cycle $i$, then we say that a \textit{synthesis defect occurs at cycle $i$}.
Formally, we have the following definition.

\begin{definition}\label{def:syndef}
    Suppose that $\sC=(\bfc_1,\bfc_2,\ldots,\bfc_M)\in(\Sigma^n)^M$. For $\Delta\subseteq[4n]$, we say that $\sC$ suffers from \textit{synthesis defects} at the cycles in $\Delta$ if for $i\in [M]$, the symbols of $\bfc_i$ synthesized at cycles $\Delta$ are deleted and results in $\syndef_{{\Delta}}(\bfc_i)$.
    Therefore, we define $\syndef_\Delta(\sC)\triangleq(\syndef_{{\Delta}}(\bfc_1),\ldots,\syndef_{{\Delta}}(\bfc_M))$.
\end{definition}

For simplicity, we assume the strands are \textit{ordered} and so, the identities of the strands are known to both the sender and receiver. In practice, this means that the strands are assigned unique indices and we assume that the indices are received error-free.

\begin{example}\label{exa:1}
  Let $M=3$ and $n=5$. Consider the set of strands $\sC=(\bfc_1,\bfc_2,\bfc_3)$, where
  \begin{align*}
    \bfc_1 & = 31411 & (~\cycle(\bfc_1)&=(3,5,8,9,13)~),\\
    \bfc_2 & = 12213 & (~\cycle(\bfc_2)&=(1,2,6,9,11)~),\\
    \bfc_3 & = 14131 & (~\cycle(\bfc_3)&=(1,4,5,7,9)~)\,.
  \end{align*}
  When $\Delta=\{1\}$, we have that
  \[\syndef_{\Delta}(\sC)=(31411, 2213,4131)\, .\]

  On the other hand, when $\Delta=\{10\}$ or $\Delta=\{12\}$, we have
  \[\syndef_{\Delta}(\sC)=(31411, 12213,14131)\, .\]
   Observe that when synthesis defects occur in a set of strands, not all strands are erroneous. However, those strands with symbols synthesized during corresponding cycles will be deleted at those cycles. Furthermore, as illustrated in this example, it is possible that $\syndef_\Delta(\sC)=\sC$ and this poses an interesting coding challenge.\qed
\end{example}

Our goal in this paper is to study codes correcting synthesis defects. We distinguish between two cases: one where we know the defective cycles; another where the locations of defects are unknown.

\vspace{1mm}
\noindent{\em Case I}.
In the first case, even if we know the set of defective cycles, that is $\Delta$, it is possible that we cannot
infer the erroneous positions from $\syndef_{\Delta}(\sC)$.

Let us explain using the following example.
\begin{example}
  Let $M=3$ and $n=5$. Consider the set of strands $$\sC=(12341,12134,21231).$$
  When $\Delta=\{5\}$, we have that
  $$\syndef_{\{5\}}(\sC)=(1234,1234,2231).$$
  Now we want to recover $\sC$ from $\syndef_{\{5\}}(\sC)$ knowing that $\Delta=\{5\}$. For the strand $2231$, we insert a $1$ at the fifth cycle, and we get $21231$. This turns out to be the only option. 
  But for the strand $1234$, there are four possible positions to insert $1$ at the fifth cycle, which result in $11234,12134,12314,12341$. Therefore we cannot uniquely recover $\sC$ even if we know the locations of the defects.
  Furthermore, the erroneous positions corresponding to $1234$ are 2,~3,~4, and~5. 
  Hence, coding for erasures is insufficient. \qed
\end{example}
The above example gives rise to the following problem: how to recover $\sC$ with the knowledge of locations of defects?
In our study of this problem, it is enough to investigate the case where $M=1$.
In other words, we design codes for a single-strand.
This is because in solving the single-strand case, we obtain sufficient information, including the approximate deleted locations and the values of the deleted symbols, and other strands in the set cannot provide more information about the errors.

Hence, a \textit{$t$-known-synthesis-defect correcting code} allows one to uniquely recover a synthesized word, in the presence of $t$ synthesis defects when the locations of defects are known.

\begin{definition}\label{def:LocKnownSynDefCode}
  We say a code $\cC\subseteq\Sigma^n$ is a \textit{$t$-known-synthesis-defect correcting code} ($t$-KDCC) if for every pair of distinct codewords, $\bfc_1,\bfc_2\in\cC$, and for any $\Delta\subseteq[4n]$ with $|\Delta|=t$, we have that $$\syndef_{\Delta}(\bfc_1)\neq \syndef_{\Delta}(\bfc_2).$$
\end{definition}

\vspace{1mm}
\noindent{\em Case II}. On the other hand, when we do not know the locations of the defects, coding for a single strand is almost equivalent to coding for deletions. Hence, we consider $M>1$.
When the locations of the defects are unknown, we can simply employ a deletion-correcting code for each strand.
However, this incurs a high redundancy and hence, our goal is to reduce this redundancy.

Consider Example~\ref{exa:1} and the case where $\Delta=\{1\}$.
Suppose $\bfc_2=12213$ is obtained from a single-deletion-correcting code. Then when we receive $2213$, we are able to recover $\bfc_2$ and also determine $\Delta=\{1\}$.
Then we are able to use this information on $\Delta$ to correct the other strings.
Therefore, when the location of defects are unknown, our strategy is to divide the $M>1$ strands into two parts and employ two different coding schemes.
The first coding scheme not only allows us to correct the defects, but also provide approximate locations of the defects. This then allows us to employ a second coding scheme that incurs less redundancy and we describe this in detail in Section~\ref{sec:SDCC}.
Here, we formally define a \textit{$t$-synthesis-defect correcting code}.

\begin{definition}\label{def:syndefball}
 For a set of sequences $\sC=(\bfc_1,\bfc_2,\ldots,\bfc_M)\in(\Sigma^n)^M$, and $d\in[n]$, we define the \textit{synthesis defect ball} of radius $d$ of $\sC$ to be the set
 $$\ball^\syndef_d(\sC)=\{\syndef_{\Delta}(\sC):\Delta\subseteq[4n],|\Delta|\leq d\}.$$
\end{definition}

\begin{definition}\label{def:SynDefCode}
  A code $\cC\subseteq(\Sigma^n)^M$ is a \textit{$t$-synthesis-defect correcting code} ($t$-SDCC) if for every pair of distinct $\sC_1,\sC_2\in\cC$, we have
  $$\ball^{\syndef}_t(\sC_1)\cap\ball^{\syndef}_t(\sC_2)=\varnothing.$$
\end{definition}

In this paper, we provide designs for these two classes of codes.
We use redundancy to evaluate a code.
Specifically, we  define the \textit{redundancy} of a code $\cC\subseteq(\Sigma^n)^M$ to be $2Mn-\log|\cC|$.

\subsection{Organization}
 In Section \ref{sec:KDCC}, we construct $t$-KDCCs for $t=1,2$ with redundancy $\log 4$ and $\log n+18\log 3$ respectively. Specifically, we achieve this by constructing binary code capable of correcting $t$ deletions where each deletion is within a small window. When the window is large, this will be an important material for constructing SDCC. In Section \ref{sec:SDCC}, we first show that using $O(\log n)$ strands is enough to cover all the $4n$ cycles. Then, by giving single-SDCC and $2$-SDCC we show our idea for constructing $t$-SDCCs. In Section \ref{sec:Bound}, we provide the lower bounds for redundancy on $1$-KDCCs, which is at least $\log 4-o(1)$ bits of redundancy. Last, we conclude and present some future work in Section \ref{sec:future}.

\section{Constructions of Known-Synthesis-Defect Correcting Codes}\label{sec:KDCC}
In this section, we provide constructions of $t$-KDCCs for $t=1,2$. We show that knowing the locations of the defects helps us narrow each of the location of deletion into a small interval of constant length. Then, we can correct these deletions by using binary bounded-deletion-correcting codes \cite{schoeny2017}.

\begin{definition}
  For $\bfP=(P_1,P_2,\ldots,P_t)\in\mathbf{Z}^t$ we call a code \textit{$\bfP$-bounded $t$-deletion correcting code} if it can correct $t$ deletions where the $t$ deletions are located at $t$ given intervals of length $P_1,P_2,\ldots,P_t$, respectively. Furthermore, we let $P\triangleq\max\{P_1,\ldots,P_t\}$ denote the maximum $P_i$ for $1\leq i\leq t$.
\end{definition}

Our main contribution in this section is to construct $\bfP$-bounded $t$-deletion correcting codes with different ranges of $P$ for $t=1,2$. Before that, let us introduce a serial of useful lemmas to show that the knowledge of defective cycles narrows down the locations of resulting deletions.

\begin{definition}
    For $\Delta\subseteq[4n]$ we define $$B^{\Delta}(\bfx)=\{\bfy\in\Sigma^n:\syndef_\Delta(\bfx)=\syndef_\Delta(\bfy)\}$$ as all the length-$n$ words that will result in the same word after deleting the symbols located at these cycles in $\Delta$. Besides, if $|\Delta|=1$ and has only one element $\delta\in[4n]$, we also use the notation $B^{\delta}(\bfx)$ to represent the confusable ball of $\bfx$ in case of the defect of cycle $\delta$.
\end{definition}

For radius-$1$ confusable ball of $\bfx$, we have the following lemmas.

\begin{lemma}\label{lem:size of confuball}
  For $\bfx\in\Sigma^n$, and $\delta\in\cycle(\bfx)$, we have
  \begin{align*}
    &|B^{\delta}(\bfx)|=\\
    &|\{\delta-4,\delta-3,\delta-2,\delta-1\}\cap (\cycle(\syndef_{\{\delta\}}(\bfx))\cup\{0\})|.
  \end{align*}
\end{lemma}
\begin{IEEEproof}
    For every $\bfy\in B^{\delta}(\bfx)$, it has $\syndef_{\{\delta\}}(\bfy)=\syndef_{\{\delta\}}(\bfx)$ by the definition of $B^\delta(\bfx)$. Thus, we can generate $B^\delta(\bfx)$ by inserting the symbol ($\delta\bmod4$) in $\syndef_{\{\delta\}}(\bfx)$ and make sure it is at the $\delta$ cycle.

    Due to the property of the cycle sequence, any two consecutive elements in a cycle sequence have difference at most $4$. Consequently, if we want to insert a symbol making sure it is at the $\delta$ cycle, then it must be inserted after the symbol synthesized at cycle $\delta-4$ or $\delta-3$ or $\delta-2$ or $\delta-1$. So, the size of $B^{\delta}(\bfx)$ depends on $\{\delta-4,\delta-3,\delta-2,\delta-1\}$ and $\syndef_{\{\delta\}}(\bfx)$. Specially, if we insert ($\delta\bmod4$) at the beginning of $\syndef_{\{\delta\}}(\bfx)$, then we actually insert it after the cycle $0$. Hence, we conclude that $|B^{\delta}(\bfx)|=| \{\delta-4,\delta-3,\delta-2,\delta-1\}\cap (\cycle(\syndef_{\{\delta\}}(\bfx))\cup\{0\})|$.
\end{IEEEproof}

This leads to the following lemma, demonstrating the relationship between the indices of deletions in $\bfx$ and those in its confusable sequence.

\begin{lemma}\label{lem:narrow locations}
    If $\bfx\in\Sigma^{n}$ suffers from $t$ synthesis defects at cycles in $\Delta=\{\delta_1,\ldots,\delta_t\}\subseteq\cycle(\bfx)$ resulting in $\syndef_\Delta(\bfx)$, then for any $\bfy\in B^\Delta(\bfx)$ with $\cycle(\bfx)_{\{i_1,i_2,\ldots,i_t\}}=\cycle(\bfy)_{\{j_1,j_2,\ldots,j_t\}}=\Delta$, we have $|i_k-j_k|\leq 4k-1$ for $1\leq k\leq t$.
\end{lemma}
\begin{IEEEproof}
    For any $\bfy\in B^\Delta(\bfx)$, $\bfy$ can be obtained by inserting $t$ symbols successively into $\syndef_\Delta(\bfx)$ in the order $x_{i_1},x_{i_2},\ldots,x_{i_t}$ and at positions $j_1,j_2,\ldots,j_t$. By Lemma \ref{lem:size of confuball}, there are at most $4$ positions to insert $x_{i_1}$ such that it is inserted at the cycle of $\cycle(\bfx)_{i_1}$, where the index is in the range $[i_1-3,i_1+3]$, so $|i_1-j_1|\leq 3$. After inserting $x_{i_1}$ into $\syndef_\Delta(\bfx)$, each cycle of $\syndef_\Delta(\bfx)_i$ will increase by $4$ or keep unchanged for $i>i_1$. Again there are at most $4$ positions to insert $x_{i_2}$ at the cycle of $\cycle(\bfx)_{i_2}$. Since the cycles may have increased by $4$, $x_{i_2}$ is possible to be inserted near the symbol synthesized at $\cycle(\bfx)_{i_2}-4$, so $|i_2-j_2|\leq 7$. Repeating this process, we conclude that $|i_k-j_k|\leq 4k-1$ for $1\leq k\leq t$.
\end{IEEEproof}

\begin{example}
    Let $\bfx$ and $\cycle(\bfx)$ be the following, and let $\Delta=\{5,17\}$.
    \begin{table}[H]
        \setlength\tabcolsep{2pt}
        \centering
        \begin{tabular}{cccccccccccccccccccc}
        $\cycle(\bfx)$ & = & 1 & 2 & 3 & 4 & \underline{5} & 6 & 7 & 8 & 9 & 10 & 11 & 12 & 13 & 14 & 15 & 16 & \underline{17} & 18\\
        $\bfx$ & = & 1 & 2 & 3 & 4 & \underline{1} & 2 & 3 & 4 & 1 & 2 & 3 & 4 & 1 & 2 & 3 & 4 & \underline{1} & 2\vspace{-2ex}
        \end{tabular}
    \end{table}
    \noindent Then, we have $\syndef_\Delta(\bfx)\triangleq\bfx'$ as follows:
    \begin{table}[H]
        \setlength\tabcolsep{2pt}
        \centering
        \begin{tabular}{cccccccccccccccccc}
        $\cycle(\bfx')$ & = & 1 & 2 & 3 & 4 & 6 & 7 & 8 & 9 & 10 & 11 & 12 & 13 & 14 & 15 & 16 & 18\\
        $\bfx'$ & = & 1 & 2 & 3 & 4 & 2 & 3 & 4 & 1 & 2 & 3 & 4 & 1 & 2 & 3 & 4 & 2\vspace{-2ex}
        \end{tabular}
    \end{table}
    Now we want to insert two $1$s into $\syndef_\Delta(\bfx)$ such that these two $1$s are at $5$th and $17$th cycles. We first insert a $1$ at the second position of $\syndef_\Delta(\bfx)$, then we get $\bfx''$ as follows:
    \begin{table}[H]
        \setlength\tabcolsep{2pt}
        \centering
        \begin{tabular}{ccccccccccccccccccc}
        $\cycle(\bfx'')$ & = & 1 & \underline{5} & 6 & 7 & 8 & 10 & 11 & 12 & 13 & 14 & 15 & 16 & 17 & 18 & 19 & 20 & 22\\
        $\bfx''$ & = & 1 & \underline{1} & 2 & 3 & 4 & 2 & 3 & 4 & 1 & 2 & 3 & 4 & 1 & 2 & 3 & 4 & 2\vspace{-2ex}
        \end{tabular}
    \end{table}
    Next, to get a sequence in $B^\Delta(\bfx)$ we need to insert a $1$ into $\bfx''$ such it is at $17$th cycle. We insert it at the $10$th position of $\bfx''$, and we get $\bfy\in B^\Delta(\bfx)$:
    \begin{table}[H]
        \setlength\tabcolsep{2pt}
        \centering
        \begin{tabular}{cccccccccccccccccccc}
        $\cycle(\bfy)$ & = & 1 & \underline{5} & 6 & 7 & 8 & 10 & 11 & 12 & 13 & \underline{17} & 18 & 19 & 20 & 21 & 22 & 23 & 24 & 26\\
        $\bfy$ & = & 1 & \underline{1} & 2 & 3 & 4 & 2 & 3 & 4 & 1 & \underline{1} & 2 & 3 & 4 & 1 & 2 & 3 & 4 & 2\vspace{-2ex}
        \end{tabular}
    \end{table}
    Compare $\bfy$ with $\bfx$, and we have $\cycle(\bfx)_{\{5,17\}}=\cycle(\bfy)_{\{2,10\}}=\{5,17\}=\Delta$, which satisfies the conclusion of Lemma \ref{lem:narrow locations} that $|i_k-j_k|\leq 4k-1$ for $k=1,2$.\qed
\end{example}

\subsection{Reduction to binary codes}\label{subsec:reduct to binary}
In this subsection, we show that the constructions of $t$-KDCCs can be reduced to constructions of binary codes.
\begin{definition}[\hspace*{-1.2mm}\cite{tenengolts1984nonbinary}]
    Let $\bfx \in \Sigma^n$ be a quaternary word of length $n$.
    Define the {\em signature} of
    $\bfx$ to be the binary sequence $\widetilde{\bfx}$ of length $n-1$ such that
    $$\widetilde{x}_i=
    \begin{cases}
		1 &\quad \text{if } x_{i+1} \geq x_i,\\
		0 &\quad \text{if } x_{i+1} < x_i,\\
    \end{cases}$$
    for all $1 \leq i\leq n-1$.
\end{definition}

\begin{lemma}\label{lem:rec x with loc and bin}
    If $\bfx\in\Sigma^n$ suffers from $t$ synthesis defects at cycle $\Delta=\{\delta_1,\ldots,\delta_t\}\subseteq\cycle(\bfx)$ resulting in $\syndef_\Delta(\bfx)$, then with the knowledge of $\Delta$ and $\widetilde{\bfx}$, we can recover $\bfx$ from $\syndef_\Delta(\bfx)$.
\end{lemma}
\begin{IEEEproof}
    We prove this by giving an algorithm, where the inputs are $\syndef_\Delta(\bfx),\Delta=\{\delta_1,\ldots,\delta_t\}$ and $\widetilde{\bfx}$, and output is $\bfx$. First we recover the symbol synthesized at $\delta_1$ cycle. By Lemma \ref{lem:size of confuball} and \ref{lem:narrow locations}, there are at most $4$ positions to insert the symbol $(\delta_1\bmod{4})$. Specifically, it can only be inserted after symbols synthesized at cycles in $\{\delta_1-4,\delta_1-3,\delta_1-2,\delta_1-1\}$. Let $I=\{i+1:\cycle(\bfx)_{i}\in\{\delta_1-4,\delta_1-3,\delta_1-2,\delta_1-1\}\}$, then $I$ is the set of indices suitable to insert $(\delta_1\bmod{4})$. Furthermore, $|I|\leq4$ and the elements in $I$ are consecutive. Since we can know the monotonicity of $\bfx_{I}$ from $\widetilde{\bfx}$, there is only one choice in $I$ to insert $(\delta_1\bmod{4})$ keeping the monotonicity. Denote the sequence obtained by inserting $(\delta_1\bmod{4})$ into $\syndef_\Delta(\bfx)$ as $\bfx'$, and renew the inputs as $\bfx',\Delta=\{\delta_2,\ldots,\delta_t\}$ and $\widetilde{\bfx}$. After $t$ rounds of algorithm, we can get the correct sequence $\bfx$.
\end{IEEEproof}

The following is an algorithm we mentioned above to recover $\bfx$ with the knowledge of the defective cycles and the signature.

\begin{algorithm}[ht]
\SetAlgoLined
\algorithmicinput{ $\syndef_\Delta(\bfx)$, $\Delta = \{\delta_1,\ldots, \delta_t\}$ and $\widetilde{\bfx}$}\\
\algorithmicoutput{ $\bfx$}\\
Set $j=1$, $\bfx'=\syndef_\Delta(\bfx)$\;
\While{$\Delta \neq \varnothing$}{
    $I \leftarrow \{i+1 : \cycle(\bfx)_i \in \{\delta_j-4, \delta_j-3, \delta_j-2, \delta_j-1\}\}$\;
    \ForEach{$i \in I$}{
        $\bfx^i \leftarrow$ insert $(\delta_j \bmod 4)$ at position $i$ in $\bfx'$\;
        \If{$\exists~i$, such that $\widetilde{\bfx^i}_I = \widetilde{\bfx}_I$}{
            $\bfx' \leftarrow \bfx^i$\;
            Remove $\delta_j$ from $\Delta$\;
            $j=j+1$\;
        }
    }
}
$\bfx \leftarrow \bfx'$\;
\Return $\bfx$\;
\caption{Recover $\bfx$ with the knowledge of defective cycles and its signature}
\end{algorithm}

With Lemma \ref{lem:rec x with loc and bin} and the above algorithm, the problem of correcting $t$ synthesis defects with the knowledge of defective cycles reduces to recovering the signature with the information of defective cycles.

By the definition of signature, we have the following observation.
\begin{observation}[\hspace*{-1.0mm}{\cite{tenengolts1984nonbinary}}]\label{obs:index of del in signature}
For $\bfx\in\Sigma^n$, if a deletion occurs at $x_i$, then $\widetilde{x}_i$ or $\widetilde{x}_{i-1}$ will be deleted. Conversely, if the index of deletion in $\widetilde{\bfx}$ is $i$, then the index of deletion in $\bfx$ is $i$ or $i+1$.
\end{observation}
Specifically, if $x_{i-1}\leq x_i\leq x_{i+1}$ or $x_{i-1}>x_i>x_{i+1}$, then the deletion of $x_i$ will cause a deletion of $\widetilde{x}_i$. If $x_{i-1}\leq x_i> x_{i+1}$ and $x_{i-1}>x_{i+1}$, then the deletion of $x_i$ will cause a deletion of $\widetilde{x}_{i-1}$. If $x_{i-1}\leq x_i> x_{i+1}$ and $x_{i-1}\leq x_{i+1}$, then the deletion of $x_i$ will cause a deletion of $\widetilde{x}_{i}$. The case of $x_{i-1}> x_i\leq x_{i+1}$ is similar. So we have the following corollary of Lemma \ref{lem:narrow locations}.

\begin{corollary}\label{cor:narrow locations}
    Let $\bfx\in\Sigma^n$, and $\Delta\subseteq\cycle(\bfx)$ is of size $t$. For any $\bfy\in B^\Delta(\bfx)$ with $\widetilde{\bfx}_{n\setminus\{i_1,i_2,\ldots,i_t\}}=\widetilde{\bfy}_{n\setminus\{j_1,j_2,\ldots,j_t\}}$, we have $|i_k-j_k|\leq 4k$ for $1\leq k\leq t$.
\end{corollary}

With this corollary, if we want to construct a $t$-KDCC, it suffices to construct a binary $\bfP$-bounded $t$-deletion correcting code for $\bfP=(5,9,\ldots,4t+1)$.

\subsection{Constructions for $t=1$}
We first construct a single-KDCC by using binary $P$-bounded $1$-deletion correcting code with redundancy $\log 12$. Then, we show that this can be improved by giving another construction with redundancy $\log 4$.

\begin{definition}
  For $\bfx\in\ZZ^n$, the {\em VT-syndrome} of $\bfx$ is defined as $\VT(\bfx)\triangleq\sum_{i=1}^n{ix_i}$. Define $\Sum(\bfx)\triangleq\sum_{i=1}^{n}x_i$.
\end{definition}

\begin{construction}[\hspace*{-1.2mm}{\cite[Construction 1]{schoeny2017}}]
    For $a\in\ZZ_P, b\in\ZZ_2$, let the shifted Varshamov-Tenengolts code be:
    \begin{align*}
        \SVT_{a,b}(n,P)=\Big\{\bfx\in\Sigma^n:&\VT(\bfx)=a\bmod P,\\
        & \Sum(\bfx)=b\bmod2\Big\}
    \end{align*}
\end{construction}

\begin{lemma}[\hspace*{-1mm}{\cite[Lemma 4, 5]{schoeny2017}}]\label{lem:SVT}
    For $a\in\ZZ_P, b\in\ZZ_2$, the shifted VT code $\SVT_{a,b}(n,P)$ is a $P$-bounded single-deletion correcting code, and there exist $a$ and $b$ such that the redundancy is at most $\log P+1$.
\end{lemma}

\begin{construction}
    For $a\in\ZZ_5,b\in\ZZ_2$ define the code
    \begin{equation*}
        \cC_1^{KDCC}(n;a,b)=\Big\{\bfx\in\Sigma^n: \widetilde{\bfx}\in\SVT_{a,b}(n,5)\Big\}.
    \end{equation*}
\end{construction}

\begin{theorem}
    For $a\in\ZZ_5,b\in\ZZ_2$ the code $\cC_1^{KDCC}(n;a,b)$ is a $1$-KDCC, and there exist $a$ and $b$ such that the redundancy is at most $\log 10$.
\end{theorem}
\begin{IEEEproof}
    Suppose $\cC_1^{KDCC}(n;a,b)$ is not a $1$-KDCC, then there exists a $\bfy\in\cC_1^{KDCC}(n;a,b)$, such that $\bfy\in B^\delta(\bfx)$ for some $\delta\in\cycle(\bfx)$. Let $i$ and $j$ be the indices such that $\bfx_{[n]\setminus\{i\}}=\bfy_{[n]\setminus\{j\}}$, then we have $i'\in\{i-1,i\},j'\in\{j-1,j\}$ where $\widetilde{\bfx}_{[n]\setminus\{i'\}}=\widetilde{\bfy}_{[n]\setminus\{j'\}}$. By Lemma \ref{lem:narrow locations} and Corollary \ref{cor:narrow locations}, it has to be $|i-j|\leq 3$ and $|i'-j'|\leq4$ which contradict that $\widetilde{\bfx},\widetilde{\bfy}\in\SVT_{a,b}(n,5)$.

    Set $P=5$, we get that there exist $a$ and $b$ such that the redundancy of $\cC_1^{KDCC}(n;a,b)$ is at most $\log5+1=\log10$ by Lemma \ref{lem:SVT}.
\end{IEEEproof}

The above shows an example of constructing $1$-KDCC from binary $\bfP$-bounded single-deletion correcting code
In the remaining of this subsection, we construct $1$-KDCC without using the signature.

\begin{construction}
    For $a\in\ZZ_4$, define the code
    \begin{equation*}
        \cC_1^{KDCC}(n;a)=\Big\{\bfx\in\Sigma^n: \sum_{i=1}^{\lfloor\frac{n}{2}\rfloor}x_{2i}=a\bmod4\Big\}.
    \end{equation*}
\end{construction}

\begin{theorem}
    For $a\in\ZZ_4$ the code $\cC_1^{KDCC}(n;a)$ is a $1$-KDCC and there exists an $a$ such that the redundancy is at most $\log 4$ for some $a\in\ZZ_4$.
\end{theorem}
\begin{IEEEproof}
    Suppose $\cC_1^{KDCC}(n;a)$ is not a $1$-KDCC, then there exists a $\bfy\in\cC_1^{KDCC}(n;a)$, such that $\bfy\in B^\delta(\bfx)$ for some $\delta\in\cycle(\bfx)$. Again we let $i<j$ be the indices such that $\bfx_{[n]\setminus\{i\}}=\bfy_{[n]\setminus\{j\}}$. By the proof of Lemma \ref{lem:size of confuball}, if we concentrate on the different parts of $\bfx$ and $\bfy$ which are $\bfx_{[i,j]}$ and $\bfy_{[i,j]}$, then $x_i=y_j$ and $x_{k}=y_{k-1}$ for $i+1\leq k\leq j$. Besides, $x_i,\ldots,x_j$ are all different since they are synthesized within $4$ cycles. Now we discuss in three cases, $j-i=3,2,1$.
    \begin{itemize}
        \item If $j-i=1$, then we have $$\left\vert\sum_{i=1}^{\lfloor\frac{n}{2}\rfloor}x_{2i}-\sum_{i=1}^{\lfloor\frac{n}{2}\rfloor}y_{2i}\right\vert=\left\vert x_i-x_j\right\vert\in\{1,2,3\},$$ a contradiction.
        \item If $j-i=2$, then we have
        $$\left\vert\sum_{i=1}^{\lfloor\frac{n}{2}\rfloor}x_{2i}-\sum_{i=1}^{\lfloor\frac{n}{2}\rfloor}y_{2i}\right\vert= \begin{cases}
               |x_i+x_j-x_{i+1}-x_i|, \text{ if $i$ is even},\\
               |x_{i+1}-x_j|, \text{ if $i$ is odd}.
           \end{cases}$$
         Either $i$ is even or odd, we can get $\left\vert\sum_{i=1}^{\lfloor\frac{n}{2}\rfloor}x_{2i}-\sum_{i=1}^{\lfloor\frac{n}{2}\rfloor}y_{2i}\right\vert=|x_{i+1}-x_j|\in\{1,2,3\}$, a contradiction.
         \item If $j-i=3$, then $$\left\vert\sum_{i=1}^{\lfloor\frac{n}{2}\rfloor}x_{2i}-\sum_{i=1}^{\lfloor\frac{n}{2}\rfloor}y_{2i}\right\vert=|x_i+x_{i+2}-x_{i+1}-x_j|.$$ In this case, $\bfx_{[i:j]}=x_i x_{i+1} x_{i+2} x_{i+3}\in\{1234,2341,3412,4123\}$, so $|x_i+x_{i+2}-x_{i+1}-x_j|=2$, again a contradiction.
    \end{itemize}
    In all cases, $1\leq\left\vert\sum_{i=1}^{\lfloor\frac{n}{2}\rfloor}x_{2i}-\sum_{i=1}^{\lfloor\frac{n}{2}\rfloor}y_{2i}\right\vert\leq3$, which contradicts $\sum_{i=1}^{\lfloor\frac{n}{2}\rfloor}x_{2i}=\sum_{i=1}^{\lfloor\frac{n}{2}\rfloor}y_{2i}\bmod4$.

    By the pigeonhole principle, there exists $a\in \ZZ_4$, such the code $\cC_1^{KDCC}(n;a)$ has size at least $\frac{4^n}{4}$. So the redundancy is at most $\log 4$.
\end{IEEEproof}

\subsection{Construction of $t=2$}
In this subsection, we provide a $2$-KDCC by constructing a binary $\bfP$-bounded $2$-deletion correcting code for $\bfP=(P_1,P_2)$.

For $\bfx\in\{0,1\}^n$, we use $A(\bfx)$ to represent the $P\times\frac{n}{P}$ array form of $\bfx$. That is,
\begin{equation*}
  A(\bfx)=\left[\begin{array}{cccc}
            x_1 & x_{P+1} & \cdots & x_{n-P+1} \\
            x_2 & x_{P+2} & \cdots & x_{n-P+2} \\
            \vdots & \vdots & \ddots & \vdots \\
            x_P & x_{2P} & \cdots & x_n
          \end{array}\right].
\end{equation*}
Let $A(\bfx)_i$ be the $i$th row of $A(\bfx)$, where $1\leq i \leq P$.

\begin{construction}
  For $\boldsymbol a\in\ZZ_3^P$ and $b\in\ZZ_{3^Pn}$, we denote
  \begin{align*}
    \cC_2^P(n;\boldsymbol a,b)&=\Big\{\bfx\in\{0,1\}^n:  \sum_{i=1}^{P}3^{i-1}\VT(A(\bfx)_i)=b\bmod 3^Pn,\\
    & \Sum(A(\bfx)_i)=a_i\bmod 3, \text{ for } 1\leq i\leq P\Big\}.
  \end{align*}
\end{construction}

\begin{lemma}\label{lem:erasure}
  For $\boldsymbol a\in\ZZ_3^P$ and $b\in\ZZ_{3^Pn}$, the code $\cC_2^P(n;\boldsymbol a,b)$ can correct two bursts of erasures of length at most $P$.
\end{lemma}

\begin{IEEEproof}
   It is easy to see that a burst of erasures of length $P$ will cause an erasure in every row of $A(\bfx)$. If the coordinates of the two bursts of erasures overlap, or in other words, the error type is a burst of erasure of length at most $2P-1$, then we can erase more bits such that the error is two bursts of erasures of length $P$. Thus, every row of $A(\bfx)$ will has two erasures.

   Suppose the erroneous word is $\bfx'$. We can get the two  missing bits in every row by calculating $\Sum(A(\bfx)_i)-\Sum(A(\bfx')_i)$ for $1\leq i\leq P$. If the difference is $0$, then we know the two missing bits are both $0$. If the difference is $2$, then we know the two missing bits are both $1$. In these two cases, the row will be recovered correctly. If the difference is $1$, then we know the two missing bits are $1$ and $0$, but we do not know their order. In the following, we consider this case.

   Let the indices of the two erasures in $i$th row are $k_i$ and $\ell_i$, where $1\leq k_i<\ell_i\leq\frac{n}{P}$ for $1\leq i\leq P$. We have $k_i\in\{k_1,k_1-1\}$ and $\ell_i\in\{\ell_1,\ell_1-1\}$ for $2\leq i\leq P$. Suppose we have another codeword $\bfy\in\cC_2^P(n;\boldsymbol a,b)$, which results in the same word after erasing the same coordinates. Now we consider the difference of the first constraint between $\bfx$ and $\bfy$. Denote their difference as $$D\triangleq\sum_{i=1}^{P}3^{i-1}\VT(A(\bfx)_i)-\sum_{i=1}^{P}3^{i-1}\VT(A(\bfy)_i).$$ Since $\bfx\neq \bfy$, there must exist some $\{i_1,\ldots,i_p\}\subseteq[P]$, such that $A(\bfx)_{i_j} \neq A(\bfy)_{i_j}$. As $\Sum(A(\bfx)_{i_j})=\Sum(A(\bfy)_{i_j})\bmod 3$, we have $s_{i_j}=A(\bfx)_{i_j,k_{i_j}}=1-A(\bfy)_{i_j,k_{i_j}}=1-A(\bfx)_{i_j,\ell_{i_j}}=A(\bfy)_{i_j,\ell_{i_j}}$, where $s_{i_j}\in\{0,1\}$ for $1\leq j\leq p$. Therefore,
   \begin{align*}
     D & =\sum_{i=1}^{P}3^{i-1}\VT(A(\bfx)_i)-\sum_{i=1}^{P}3^{i-1}\VT(A(\bfy)_i) \\
      & =\sum_{j=1}^{p}3^{i_j-1}\VT(A(\bfx)_{i_j})-\sum_{j=1}^{p}3^{i_j-1}\VT(A(\bfy)_{i_j}) \\
      & =\sum_{j=1}^{p}3^{i_j-1}(\VT(A(\bfx)_{i_j})-\VT(A(\bfy)_{i_j}))\\
      & =\sum_{j=1}^{p}3^{i_j-1}(1-2s_{i_j})(\ell_{i_j}-k_{i_j}).
   \end{align*}
   Since $1\leq\ell_{i_j}-k_{i_j}\leq\frac{n}{P}$, we can bound the range of $D$: $$|D|\leq\sum_{j=1}^{p}3^{i_j-1}\frac{n}{P}\leq\frac{3^P-1}{2P}n,$$
   which is less than $3^Pn$.

   Now we show $D$ cannot be $0$. As $k_i\in\{k_1,k_1-1\}$ and $\ell_i\in\{\ell_1,\ell_1-1\}$ for $2\leq i\leq P$, we have $\ell_{i_j}-k_{i_j}\in\{d-1,d,d+1\}$ for $1\leq j\leq p$, where $d\triangleq\ell_{i_1}-k_{i_1}$. We claim that there exists no $1\leq u<v\leq p$ such that $\{\ell_{i_u}-k_{i_u},\ell_{i_v}-k_{i_v}\}=\{d-1,d+1\}$. If $\ell_{i_u}-k_{i_u}=d-1$, this means $\ell_{i_u}=\ell_1-1,k_{i_u}=k_1$. Since $u<v$, by the array representation it has $\ell_{i_v}\leq\ell_{i_u}$ and $k_{i_v}\leq k_{i_u}$. The possibilities of $(\ell_{i_v},k_{i_v})$ are $(\ell_1-1,k_1)$ and $(\ell_1-1,k_1-1)$. Therefore $\ell_{i_v}-k_{i_v}\in\{d-1,d\}$. If $\ell_{i_u}-k_{i_u}=d+1$, this means $\ell_{i_u}=\ell_1,k_{i_u}=k_1-1$. Similarly we can get $\ell_{i_v}-k_{i_v}\in\{d,d+1\}$. The above claim shows that if there exist a least index $1\leq j\leq p$ such that $\ell_{i_j}-k_{i_j}=d'\neq d$, then for all $j<j'\leq p$,  $\ell_{i_{j'}}-k_{i_{j'}}\in\{d',d\}$. In other words, $\ell_{i_j}-k_{i_j}\in\{d-1,d\}$ or $\ell_{i_j}-k_{i_j}\in\{d,d+1\}$ for $1\leq j\leq p$.
   \begin{itemize}
     \item If $s_{i_p}=0$, then
   \begin{align*}
     D & = \sum_{j=1}^{p}3^{i_j-1}(1-2s_{i_j})(\ell_{i_j}-k_{i_j})\\
      & \geq 3^{i_p-1}(\ell_{i_p}-k_{i_p})-\sum_{j=1}^{p-1}3^{i_j-1}(\ell_{i_j}-k_{i_j})\\
      & \geq 3^{i_p-1}(\ell_{i_p}-k_{i_p})-\sum_{j=1}^{p-1}3^{i_j-1}(\ell_{i_p}-k_{i_p}+1)\\
      & \geq 3^{i_p-1}(\ell_{i_p}-k_{i_p})-\frac{3^{i_p-1}-1}{2}(\ell_{i_p}-k_{i_p}+1)\\
      & =\frac{3^{i_p-1}+1}{2}(\ell_{i_p}-k_{i_p})-\frac{3^{i_p-1}-1}{2}\\
      & \geq \frac{3^{i_p-1}+1}{2}-\frac{3^{i_p-1}-1}{2}\\
      & = 1.
   \end{align*}
     \item If $s_{i_p}=1$, then
   \begin{align*}
     D & = \sum_{j=1}^{p}3^{i_j-1}(1-2s_{i_j})(\ell_{i_j}-k_{i_j})\\
      & \leq \sum_{j=1}^{p-1}3^{i_j-1}(\ell_{i_j}-k_{i_j})-3^{i_p-1}(\ell_{i_p}-k_{i_p})\\
      & \leq \sum_{j=1}^{p-1}3^{i_j-1}(\ell_{i_p}-k_{i_p}+1)-3^{i_p-1}(\ell_{i_p}-k_{i_p})\\
      & \leq \frac{3^{i_p-1}-1}{2}(\ell_{i_p}-k_{i_p}+1)-3^{i_p-1}(\ell_{i_p}-k_{i_p})\\
      & =-\frac{3^{i_p-1}+1}{2}(\ell_{i_p}-k_{i_p})+\frac{3^{i_p-1}-1}{2}\\
      & \leq -\frac{3^{i_p-1}+1}{2}+\frac{3^{i_p-1}-1}{2}\\
      & = -1.
   \end{align*}
   \end{itemize}
   In either case, $D$ cannot be $0$, which contradicts the fact that $\bfx,\bfy\in\cC_2^P(n;\bfa,b)$.
\end{IEEEproof}

\begin{theorem}
  For $\bfP=(P_1,P_2)$,  $\boldsymbol a\in\ZZ_3^P$ and $b\in\ZZ_{3^Pn}$, the code $\cC_2^P(n;\boldsymbol a,b)$ is a binary $\bfP$-bounded two-deletion correcting code, and there exist $\bfa$ and $b$ such that it has redundancy at most $\log n+2P\log 3$.
\end{theorem}
\begin{IEEEproof}
  If we know the knowledge of the two intervals containing deletions, then the bits outside these two intervals are error-free and can be put in their correct positions. For example, if $\bfx$ deletes two bits in two intervals $[i,i+P_1-1]$ and $[j,j+P_2-1]$ respectively, where $1\leq i\leq j\leq n-P+1$. Let $\bfx'$ denote the erroneous word of $\bfx$, then by setting
  \begin{align*}
      \bfx''=~ & x'_1,\ldots,x'_{i-1},\underbrace{?,\ldots,?}_{P_1},x'_{i+P_1-1},\ldots,\\
              & x'_{j-2},\underbrace{?,\ldots,?}_{P_2},x'_{j+P_2-2},\ldots,x'_{n-2},
  \end{align*}
  we get a word resulted from $\bfx$ by two bursts of erasures of length $P_1,P_2$. By Lemma \ref{lem:erasure}, we can recover $\bfx$ from $\bfx''$. So the code $\cC_2^P(n;\boldsymbol a,b)$ can always correct two deletions if we have the information of two intervals containing deletions.

  By the pigeonhole principle, there exists $\boldsymbol a\in \ZZ_3^P, b\in\ZZ_{3^Pn}$, such the code $\cC_2^P(n;\boldsymbol a,b)$ has size at least $$\frac{2^n}{3^Pn\cdot 3^P}.$$ So the redundancy is at most $\log n+2P\log3$.
\end{IEEEproof}

Now we are ready to give the construction of $2$-KDCC. By Corollary \ref{cor:narrow locations}, we can set $\bfP=(5,9)$, thus $P=\max\{5,9\}=9$.

\begin{construction}
    For $\boldsymbol a\in\ZZ_3^9$, $b\in\ZZ_{3^9n}$ and $P=9$, we denote
    $$\cC_2^{KDCC}(n;\bfa,b)=\Big\{\bfx\in\Sigma^n: \widetilde{\bfx}\in\cC_2^P(n;\bfa,b)\Big\}.$$
\end{construction}

\begin{theorem}
    For $\boldsymbol a\in\ZZ_3^9$ and $b\in\ZZ_{3^9n}$, the code $\cC_2^{KDCC}(n;\bfa,b)$ is a $2$-KDCC, and there exists parameters such that the redundancy is at most $\log n+18\log 3$.
\end{theorem}
\begin{IEEEproof}
    By Lemma \ref{lem:rec x with loc and bin}, $\cC_2^{KDCC}(n;\bfa,b)$ is a $2$-KDCC, as we can uniquely recovery each signature of the codeword which is in a $(5,9)$-bounded $2$-deletion correcting code. Furthermore, there exists choices of $\boldsymbol a\in\ZZ_3^9$ and $b\in\ZZ_{3^9n}$, such that the redundancy of $\cC_2^{KDCC}(n;\bfa,b)$ is at most $2n-\log\frac{4^n}{3^{18}n}=\log n+18\log3$.
\end{IEEEproof}

We have constructed a binary $\bfP$-bounded $2$-deletion correcting code in this subsection. But if we let $P=c\log n$ for some $c>1$, then the redundancy of $\cC_2^P(n;\bfa,b)$ will be at least $(2\log 3+1)\log n$. To decrease the redundancy when $P$ is large, we further construct another binary $\bfP$-bounded $2$-deletion correcting code with redundancy $2\log n+o(\log n)$ even when $P=O(\log n)$. Although we do not require $P$ to be large in this section, our $\bfP$-bounded $2$-deletion correcting code for $P=O(\log n)$ will be useful in our construction of $2$-SDCC.

The following lemma will lead to a $\bfP$-bounded $2$-deletion correcting code, and the proof can be found in Appendix \ref{sec:large P bounded}.
\begin{lemma}\label{lem:bounded two del}
    For any integers $P_1,P_2\geq2$ and $n\geq3$, there exists a function $\cE:\{0,1\}^n\rightarrow\{0,1\}^{n+r(n,P_1,P_2)}$, computable in linear time, such that for any $\bfx\in\{0,1\}^n$, if $\cE(\bfx)$ suffers two deletions at two intervals of lengths $P_1,P_2$ resulting in $\cE'(\bfx)$, then given the information of these two intervals and $\cE'(\bfx)$ we can uniquely recover $\bfx$, where $i,j\in[n+r(n,P_1,P_2)]$ are the indices of deletions, and $r(n,P_1,P_2)=2\log n+7\log\log n+14\log(P_1+P_2)+\log P+O(\log\log(P_1+P_2))$.
\end{lemma}

Let $\cC_2^\bfP=\{\bfx\in\{0,1\}^n: \bfx=\cE(\bfz)_{[k]}01\cE(\bfz)_{[k+1,n-2]}, \bfz\in\{0,1\}^k\}$, where $k+r(k,P_1,P_2)=n-2$. We have the following theorem.

\begin{theorem}
   $\cC_2^\bfP$ is a binary $\bfP$-bounded two-deletion correcting code of redundancy $2\log n+7\log\log n+14\log(P_1+P_2)+\log P+O(\log\log(P_1+P_2))$.
\end{theorem}

\section{Constructions of Synthesis-Defect Correcting Codes}\label{sec:SDCC}

In this section, we will provide $t$-SDCC for $t=1,2$. First we show that we can localize each defect to a window of length $O(\log n)$ by letting the signatures of $O(\log n)$ strands belong to binary $t$-deletion correcting codes and with some period constraints. Then, it suffices to employ another coding scheme which incurs less redundancy so that we can correct the remaining strands.

\subsection{Defect-Locating Strands}

In this subsection, we code for a set of ordered sequences. We show that using $O(\log n)$ strands is enough to cover all the $4n$ cycles, so we do not need to let all $M$ strands belong to deletion-correcting code. By coding these $O(\log n)$ strands, we can localize each defect to a window. With this, we are prepared to provide the construction of $t$-SDCCs for $t=1,2$.

\begin{definition}\label{def:shift}
        Define a \textit{shift} operation on $\bfx\in\Sigma^n$ with parameter $1-\cycle(\bfx)_1\le a\le 4n-\cycle(\bfx)_n$ as $S_a(\bfx) = (S_a(\bfx)_1,\ldots,S_a(\bfx)_n)$, where $\cycle(S_a(\bfx))=(\cycle(\bfx)_1+a,\ldots,\cycle(\bfx)_n+a)$.
\end{definition}

In other word, the shift operation on $\bfx$ increases each cycle of $\bfx$ by $a$. Here, we suppose the sequence $S_a(\bfx)$ can be synthesized from the $\cycle(\bfx)_1+a$ cycle and each sequence has a parameter $a$ that will be known only by the synthesis machine.

\begin{example}\label{exa:shift}
    Let $\bfx=(1,3,2,1,1,4,3,2,3,4)$, then we know that $\cycle(\bfx)=(1,3,6,9,13,16,19,22,23,24)$.
    Let $a=1$, then $S_1(\bfx)=(2,4,3,2,2,1,4,3,4,1)$, and $\cycle(S_1(\bfx))=(2,4,7,10,14,17,20,23,24,25)$.
    Let $a=6$, then $S_1(\bfx)=(3,1,4,3,3,2,1,4,1,2)$, and $\cycle(S_1(\bfx))=(7,9,12,15,19,22,25,28,29,30)$.
\end{example}

\begin{lemma}\label{lem:cover1}
    For $\bfx\in\Sigma^n$, we have $\cycle(S_t(\bfx)) \cup \cycle(S_{t+1}(\bfx)) \cup \cycle(S_{t+2}(\bfx)) \cup \cycle(S_{t+3}(\bfx))=\{\cycle(\bfx)_1+t,\ldots,\cycle(\bfx)_n+t+3\}$, for $1-\cycle(\bfx)_1\leq t\leq4n-\cycle(\bfx)_n-3$.
\end{lemma}

\begin{IEEEproof}
    By the definition of shift, $\cycle(S_t(\bfx))=(\cycle(\bfx)_1+t,\ldots,\cycle(\bfx)_n+t)$. Thus we have $$\cup_{a=t}^{t+3}\cycle(S_a(\bfx))_i=\{\cycle(\bfx)_i+t,\ldots,\cycle(\bfx)_i+t+3\},$$
    for $i\in[n]$.
    As the two elements in a cycle sequence have distance at most $4$, we have $\cycle(S_{t+3}(\bfx))_i=\cycle(\bfx)_i+t+3\geq \cycle(\bfx)_{i+1}+t-1=\cycle(S_t(\bfx))_{i+1}-1$, which means $(\cup_{a=t}^{t+3}\cycle(S_a(\bfx))_i)\bigcup(\cup_{a=t}^{t+3}\cycle(S_a(\bfx))_{i+1})=\{\cycle(\bfx)_i+t,\ldots,\cycle(\bfx)_{i+1}+t+3\}$. Therefore, we have
    $$\bigcup_{i=1}^n\cup_{a=t}^{t+3}\cycle(S_a(\bfx))_i=\{\cycle(\bfx)_1+t,\ldots,\cycle(\bfx)_n+t+3\},$$
    which complete the proof.
\end{IEEEproof}

\begin{lemma}\label{lem:cover2}
    For $\bfc_1,\ldots,\bfc_{M_1}\in\sC$ and $t\in[1,3n+1]$, there always exist $a_1,\ldots,a_{M_1}$, such that $$[t,n+t-1]\subseteq\bigcup_{i=1}^{M_1}\cycle(S_{a_i}(\bfc_i)),$$ where $M_1=\frac{1}{2-\log 3}\log n+1$, $a_i\in[-3,t+2]\cap[1-\cycle(\bfc_i)_1,4n-\cycle(\bfc_i)_n]$, and $1\leq t\leq 3n+1$.
\end{lemma}
\begin{IEEEproof}
    First of all, we choose $a_1=t-\cycle(\bfc_1)_1$, so $\cycle(S_{a_1}(\bfc_1))=\{t,t+\cycle(\bfc_1)_2-\cycle(\bfc_1)_1,\ldots,t+\cycle(\bfc_1)_n-\cycle(\bfc_1)_1\}$. Denote $T_1=[t,n+t-1]\setminus\cycle(S_{a_1}(\bfc_1))$. Since every two consecutive elements in $\cycle(S_{a_1}(\bfc_1))$ have difference at most 4, we have that $$|T_1|\leq n-\frac{1}{4}n=\frac{3}{4}n.$$

    Then, we choose $a_2$. By Lemma \ref{lem:cover1}, $\cycle(S_{t-\cycle(\bfc_2)_1}(\bfc_2))\cup\cycle(S_{t-\cycle(\bfc_2)_1+1}(\bfc_2))\cup\cycle(S_{t-\cycle(\bfc_2)_1+2}(\bfc_2))\cup\cycle(S_{t-\cycle(\bfc_2)_1+3}(\bfc_2))=\{t,\ldots,\cycle(\bfc_2)_n-\cycle(\bfc_2)_1+t+3\}\supset[t,n+t-1]$. We claim that for $a_2\in\{t-\cycle(\bfc_2)_1,t-\cycle(\bfc_2)_1+1,t-\cycle(\bfc_2)_1+2,t-\cycle(\bfc_2)_1+3\}$, there always exists a choice, such that $|T_1\cap\cycle(S_{a_2}(\bfc_2))|\geq\frac{1}{4}|T_1|$. Otherwise, $|T_1\cap\cycle(S_{a_2}(\bfc_2))|<\frac{1}{4}|T_1|$ for any $a_2\in\{t-\cycle(\bfc_2)_1,t-\cycle(\bfc_2)_1+1,t-\cycle(\bfc_2)_1+2,t-\cycle(\bfc_2)_1+3\}$. Then, we have $|T_1|=|T_1\cap[t,n+t-1]|=|T_1\cap\bigcup_{i=0}^3(\cycle(S_{t-\cycle(\bfc_2)_1+i}(\bfc_2)))|=|\bigcup_{i=0}^{3}(T_1\cap\cycle(S_{t-\cycle(\bfc_2)_1+i}(\bfc_2)))|\leq\sum_{i=0}^3|T_1\cap\cycle(S_{t-\cycle(\bfc_2)_1+i}(\bfc_2))|<|T_1|$, which is a contradiction. Let $a_2$ be the one such that $|T_1\cap\cycle(S_{a_2}(\bfc_2))|\geq\frac{1}{4}|T_1|$. Denote $T_2=[t,n+t-1]\setminus(\cycle(S_{a_1}(\bfc_1))\cup\cycle(S_{a_2}(\bfc_1)))$, then we have $$|T_2|=|T_1\setminus(T_1\cap\cycle(S_{a_2}(\bfc_2)))|\leq\frac{3}{4}|T_1|\leq\frac{9}{16}n.$$

    Repeating this, we get $|T_{M_1-1}|\leq(\frac{3}{4})^{M_1-1}n=1$. Same as previous, we choose $a_{M_1}$, such that $|T_{M_1-1}\cap\cycle(S_{a_{M_1}}(\bfc_{M_1}))|\geq1$. So when we finish choosing $a_1,\ldots,a_{M_1}$, we can make sure $[t,n+t-1]$ is covered by $\cycle(S_{a_i}(\bfc_i))$ for $i\in[M_1]$.
\end{IEEEproof}

\begin{corollary}\label{cor:cover}
    For $\bfc_1,\ldots,\bfc_{4M_1}\in\sC$, there always exist $a_1,\ldots,a_{4M_1}\in[-3,3n+1]$, such that $[1,4n]\subseteq\bigcup_{i=1}^{4M_1}\cycle(S_{a_i}(\bfc_i))$, where $M_1=\frac{1}{2-\log 3}\log n+1$.
\end{corollary}

Lemma \ref{lem:cover2} and Corollary \ref{cor:cover} show that we can use $O(\log n)$ strands to cover all synthesis cycles. So, we can know the approximate locations of all the defects if these $O(\log n)$ strands are corrected. To correct the quaternary strands, our solution in this paper it to correct their signatures first. Then, the locations of deletions in signatures will provide us the information of defects in quaternary strands.

\begin{lemma}\label{lem:range of defects}
    If $\bfx\in\Sigma^n$ suffers from a synthesis defect at cycle $\delta_1\in[4n]$ and result in $\bfx'$, then for any $\bfy\in\Sigma^n$ with $\bfy_{[n]\setminus\{j\}}=\bfx'$ and $\widetilde{\bfy}=\widetilde{\bfx}$, we have $|\delta_2-\delta_1|\leq 4P+4$, where $P$ is the length of the longest run in $\widetilde{\bfx}$ and $j\in[n]$ is the index satisfying $\cycle(\bfy)_j=\delta_2$.
\end{lemma}
\begin{IEEEproof}
    Suppose the synthesis defect causes a deletion at $i$th position of $\bfx$, then by Observation \ref{obs:index of del in signature} the location of deletion in $\widetilde{\bfx}$ is $i'\in\{i-1,i\}$. Since $\bfy_{[n]\setminus\{j\}}=\bfx_{[n]\setminus\{i\}}$, $\bfy$ can be obtained by inserting $y_j$ in $\bfx'$. Besides, due to $\widetilde{\bfy}=\widetilde{\bfx}$, the insertion of $y_j$ must lead to an insertion at the same run with $\widetilde{x}_{i'}$. So, if the insertion of $y_j$ causes an insertion at $j'$th position of $\widetilde{\bfx'}$, then we have $|j'-i'|\leq P-1$. As $j'\in\{j-1,j\}$, we have $|j-i|\leq P.$
    Without loss of generality, we assume $j>i$. Furthermore, $\cycle(\bfx)_{i+1}\in[\delta_1+1,\delta_1+4]$, so we have $\cycle(\bfy)_{j-1}=\cycle(\bfx')_{j-1}\leq\cycle(\bfx)_j\leq\delta_1+4(j-i)$, and thus $\cycle(\bfy)_j\leq\cycle(\bfy)_{j-1}+4\leq\delta_1+4(j-i)+4\leq\delta_1+4P+4$. Therefore, we have $|\delta_2-\delta_1|\leq 4P+4$.
\end{IEEEproof}

Notice that, even if the signature of a strand is recovered from $t$ deletions, the locations of errors may not be determined exactly. For example, if a strand $\bfx\in\Sigma^n$ suffers from $t$ deletions, which causes $t$ consecutive deletions in $\widetilde{\bfx}$, where $\widetilde{x}_i=\widetilde{x}_{i+t}$ for $1\leq i\leq n-t-1$, then we cannot know the locations of deletions since deleting any $t$ consecutive bits from $\widetilde{\bfx}$ will result in the same one. Intuitively, if we want to locate each of the location of error in a signature uniquely, then we need to make sure $\widetilde{\bfx}$ cannot have period $p$ for all $1\leq p\leq t$, which is a strict condition. In the following, we use the regular sequence from \cite[Definition 3]{guruswami2021} to handle the cases of $t=1,2$.

\begin{definition}[{\hspace*{-1.5mm}\cite[Definition 3]{guruswami2021}}]
    A sequence $\bfx\in\{0,1\}^n$ is \textit{regular} if each consecutive substring of $\bfx$ of length $d\log n$ contains both $00$ and $11$.
\end{definition}

\begin{lemma}[{\hspace*{-1.5mm}\cite[Lemma 11]{guruswami2021}}]
    Denote $\cR(n)$ the set of all regular binary sequences of length $n$ with $d=7$, then $|\cR(n)|\geq 2^{n-1}$.
\end{lemma}

\subsection{$1$-Synthesis-Defect Correcting Code}

Next we are prepared to provide a construction of $1$-SDCC, which shows our general strategy to construct synthesis-defect correcting code. Before that, let us introduce the well-known \textit{Varshamov-Tenengolts (VT) codes}.

\begin{lemma}[\hspace*{-2mm}\cite{varshamov1965code}]
    The Varshamov-Tenengolts (VT) code $\VT(n;a)=\{\bfx\in\{0,1\}^n:\VT(\bfx)=a\bmod(n+1)\}$ is a single-deletion correcting code, and there exists $a\in\ZZ_{n+1}$ such that the redundancy of $\VT(n;a)$ is at most $\log(n+1)$.
\end{lemma}

\begin{construction}\label{con:1SDCC}
    For $\bfs\in\ZZ_4^{4M_1}$, $\bfb\in\ZZ_{n+1}^{4M_1}$, $\bfd\in\ZZ_{P+1}^{M-4M_1}$ and $\bfe\in\ZZ_2^{M-4M_1}$, where $P=28\log n+5$ and $M_1=\frac{1}{2-\log 3}\log n+1$, we denote
    \begin{align*}
        \cC_1^{SD}=\Big\{&\sC\in(\Sigma^n)^M: \bfc_i=S_{a_i}(\bfx_i), \Sum(\bfx_i)=s_i\bmod4,\\
        &\text{ and }\widetilde{\bfx_i}\in \VT(n;b_i)\cap\cR(n), \text{ for }i\in[4M_1],\\
        &\widetilde{\bfc_i}\in\SVT_{d_{i-4M_1},e_{i-4M_1}}(n,P+1) \text{ for }i\in[4M_1+1,M]\Big\},
    \end{align*}
     where $a_i$ is chosen based on Corollary \ref{cor:cover} for $1\leq i\leq 4M_1$.
\end{construction}

\begin{theorem}\label{thm:1sdcc}
    For $\bfs\in\ZZ_4^{4M_1}$, $\bfb\in\ZZ_{n+1}^{4M_1}$, $\bfd\in\ZZ_{P+1}^{M-4M_1}$ and $\bfe\in\ZZ_2^{M-4M_1}$, where $P=28\log n+5$, the code $\cC_1^{SD}$ is a $1$-SDCC, and there exist choices of parameters, such that the redundancy of $\cC_1^{SD}$ is at most $\frac{4}{2-\log 3}(\log n)^2+M(\log\log n+6)-\Omega(\log n\log\log n)$.
\end{theorem}

\begin{IEEEproof}
    By Corollary \ref{cor:cover}, we have $\cup_{i=1}^{4M_1}\cycle(\bfc_i)=[4n]$. So if there is a defective cycle, then there must be a $\bfc_i$ suffering from a deletion for $i\in[4M_1]$. As $\widetilde{\bfx_i}$ belongs to a VT code, we can recover $\widetilde{\bfx_i}$ from a deletion uniquely. Besides, we know $\Sum(\bfx_i)\bmod4$ and $\Sum(\bfx'_i)\bmod 4$ for the erroneous sequence $\bfx'$, so we can calculate the value of deleted symbol. With the symbol and the signature, we can recover $\bfx$ in the same way as non-binary VT code \cite{tenengolts1984nonbinary}.

    Due to $\widetilde{\bfx_i}\in\cR(n)$, the length of a run in $\widetilde{\bfx_i}$ is at most $7\log n$. By Lemma \ref{lem:range of defects}, we know the defective cycle is within an interval of length $P=28\log n+5$.

    Now we show how to recover $\bfc_i$ for $i\in[4M_1+1,M]$. In the previous step, we have already located the defective cycle at a length-$P$ interval, so the deletions in the remaining $\widetilde{\bfc_i}$s are within an interval of length at most $P+1$. Since $\widetilde{\bfc_i}$ is in a shifted VT code $\SVT_{d_{i-4M_1},e_{i-4M_1}}(n,P+1)$, we can recover it. As we have already known the value of defective cycle by recovering $\bfc_i$ for $i\in[4M_1]$, we can recover $\bfc_i$ from $\widetilde{\bfc_i}$ for $i\in[4M_1+1,M]$.

    Now we calculate the redundancy of $\cC_1^{SD}$. There are $4M_1$ regular sequences belonging to VT code, and the sum of symbols in each sequence is equal to some value modular $4$. So there exists $\bfs\in\ZZ_4^{4M_1}$, $\bfb\in\ZZ_{n+1}^{4M_1}$, such the redundancy for this part is at most $4M_1\log(n+1)+12M_1$. For the remaining sequences, the redundancy for their part is at most $(M-4M_1)\log2(P+1)$. So the total redundancy is at most
    $$\frac{4}{2-\log 3}(\log n)^2+M(\log\log n+6)-\Omega(\log n\log\log n).$$
\end{IEEEproof}

The above $1$-SDCC $\cC_1^{SD}$ shows a high-level idea to construct $t$-SDCC: use $O(\log n)$ strands cover all cycles and let them belong to a $t$-deletion correcting code with some period constraints, then let the remaining sequences belong to some bounded deletion-correcting codes.

\subsection{$2$-Synthesis-Defect Correcting Code}

In the case of $t=1$, each strand either suffers from $1$ deletion or keeps unchanged. However, if $t\geq 2$, each strand will suffer $t'$ deletions for $t'\in[0,t]$. Even when we locate these $t$ deletions to $t$ interval, we do not know each of the remaining $M-4M_1$ strands suffers from which of these $t$ deletions. So we cannot locate the approximate locations of $t'$ deletions for the remaining $M-4M_1$ strands. Nevertheless, we can solve the case for $t=2$, which is more complicated than the case $t=1$.

We first present a quaternary two-deletion correcting code, which is constructed by using the signature.

\begin{lemma}[{\hspace*{-1.2mm}\cite[Theorem 7 and 8]{guruswami2021}}]
    There is a binary code $\cC_2\subseteq\{0,1\}^n$ with each codeword is regular and capable of correcting two deletions. The redundancy of $\cC_2$ is at most $4\log n+10\log\log n+O(1).$
\end{lemma}

\begin{definition}
    Define the run sequence of $\bfx\in\{0,1\}^n$ as $R(\bfx)$, where $R(\bfx)_1=1$, and for $1\leq i\leq n-1$, $$R(\bfx)_{i+1}=\begin{cases}
        R(\bfx)_{i}+1, &\text{ if } x_{i+1}\neq x_i,\\
        R(\bfx)_{i}, &\text{ if } x_{i+1}=x_i.
    \end{cases}$$
    Furthermore, let $r(\bfx)=R(\bfx)_n$ be the number of run in $\bfx$.
\end{definition}

For example, if $\bfx=10010111$, then $R(\bfx)=12234555$. The following lemma comes from \cite{liu2024explicit}, but complements their flaw.

\begin{lemma}[{\hspace*{-1.2mm}\cite[Lemma III.1]{liu2024explicit}}]\label{lem:liu2024}
    For any $\bfx\in\Sigma^n$, given the values $\sum_{i=1}^{r(\widetilde{\bfx})}x_i\cdot R(\widetilde{\bfx})_i(\bmod~4n)$ and $\widetilde{\bfx}$, if $\bfx$ suffers from two deletions resulting in $\bfx'$, and the corresponding two deleted bits in $\widetilde{\bfx}$ are not consecutive and in a period-$2$ alternate substring of length greater than $2$, then we can uniquely recovery $\bfx$ with the two values of deleted symbol.
\end{lemma}

With this lemma, it suffices to consider the case where two deletions in $\widetilde{\bfx}$ are consecutive and in a period-$2$ alternate substring. In this case, deleting any two consecutive bits in a period-$2$ alternate substring will result in the same string, so we cannot obtain the runs where the two deletions are deleted from.

\begin{lemma}\label{lem:del01}
    For $\bfx\in\Sigma^n$, if two deletions of $\bfx$ cause two consecutive deletions in an alternate substring of $\widetilde{\bfx}$, then the two deletions in $\bfx$ are consecutive or separated by one position.
\end{lemma}

\begin{IEEEproof}
    By Observation \ref{obs:index of del in signature}, a deletion indexed at $i$ in $\widetilde{\bfx}$ implies the deletion in $\bfx$ is indexed at $i$ or $i+1$. Suppose two consecutive deletions in $\widetilde{\bfx}$ caused by the two deletions in $\bfx$ have indices $i$ and $i+1$ for $1\leq i\leq n-1$, then it follows that the two deletions in $\bfx$ have indices $i,i+1$ or $i,i+2$ or $i+1,i+2$, which completes the proof.
\end{IEEEproof}

\begin{definition}
    For $\bfx\in\Sigma^n$ and $\sigma\in\Sigma$, define $N_\sigma(\bfx)$ the number of $\sigma$ in $\bfx$. Furthermore, let $\bfx^\sigma\in[n]^{N_\sigma(\bfx)}$ denote the sequence consists of the indices of $\sigma$.
\end{definition}

For example, if $\bfx=122124123$, then $N_1(\bfx)=3,N_2(\bfx)=4,N_3(\bfx)=1,N_4(\bfx)=1$ and $\bfx^1=147,\bfx^2=2358,\bfx^3=9,\bfx^4=6$. Now we present a lemma which shows that the quaternary sequence is unique if it consists of two symbols of given numbers, and has an alternate signature.

\begin{lemma}\label{lem:unique quaternary}
    If $\bfx,\bfy\in\Sigma^n$ satisfying that $N_{\sigma_1}(\bfx)=N_{\sigma_1}(\bfy)>0$, $N_{\sigma_2}(\bfx)=N_{\sigma_2}(\bfy)>0$, $N_{\sigma_3}(\bfx)=N_{\sigma_3}(\bfy)=N_{\sigma_4}(\bfx)=N_{\sigma_4}(\bfy)=0$ for $\Sigma=\{\sigma_1,\sigma_2,\sigma_3,\sigma_4\}$, and $\widetilde{\bfx}=\widetilde{\bfy}=0101\ldots$ or $1010\ldots$, then we have $\bfx=\bfy$.
\end{lemma}

\begin{IEEEproof}
    Without loss of generality, we suppose $\sigma_1<\sigma_2$. We prove the case $\widetilde{\bfx}=\widetilde{\bfy}=0101\ldots$, the other can be proved similarly. For every $\widetilde{x}_i\widetilde{x}_{i+1}=01$, it must have $x_i x_{i+1}=\sigma_2\sigma_1$ due to $\widetilde{x}_i=0$. Thus, if $n$ is even, then we have $\bfx=\bfy=(\sigma_2\sigma_1)^{n/2}$. If $n$ is odd, we can determine all symbols except the last one. As $\widetilde{x}_{n-1}=1$ and $x_{n-1}=\sigma_1$, the last symbol $x_n$ could be either $\sigma_1$ or $\sigma_2$. For the reason that $N_{\sigma_1}(\bfx)=N_{\sigma_1}(\bfy)$ and $N_{\sigma_2}(\bfx)=N_{\sigma_2}(\bfy)$, we must have $\bfx=\bfy$, which completes the proof.
\end{IEEEproof}

\begin{construction}
    For $\bfa\in\ZZ_3^4,\bfb\in\ZZ_{14\log n}^4$ and $c\in\ZZ_{nq}$, denote the code
    \begin{align*}
        &\cC_2^{D}=\Big\{\bfx\in\Sigma^n: \widetilde{\bfx}\in\cC_2,\sum_{i=1}^{r(\widetilde{\bfx})}x_i\cdot R(\widetilde{\bfx})_i=c \bmod{4n}, \text{ and}\\
        &N_\sigma(\bfx)=a_\sigma\bmod3, \Sum(\bfx^\sigma)=b_\sigma\bmod{14\log n}\text{ for }\sigma\in\Sigma\Big\}.
    \end{align*}
\end{construction}

\begin{theorem}
    For $\bfa\in\ZZ_3^4,\bfb\in\ZZ_{14\log n}^4$ and $c\in\ZZ_{nq}$, the code $\cC_2^D$ is a quaternary two-deletion correcting code, and there are parameters such the redundancy of $\cC_2^D$ is at most $5\log n+14\log\log n+O(1)$.
\end{theorem}

\begin{IEEEproof}
    For $\bfx\in\cC_2^D$, denote the erroneous sequence of $\bfx$ as $\bfx'$. Since the signature $\widetilde{\bfx}$ belongs to a binary two-deletion correcting code $\cC_2$, so we can recovery $\widetilde{\bfx}$ from $\widetilde{\bfx'}$. If the two deletions in $\widetilde{\bfx}$ are not consecutive or are consecutive $00$ or $11$, then by Lemma \ref{lem:liu2024} we can recovery $\bfx$ uniquely.

    On the other hand, if the two deletions in $\widetilde{\bfx}$ are consecutive $01$ or $10$ in an alternate substring of length greater than $2$, then we can make sure the locations of these two deletions are in an interval of length $7\log n$ for the reason that $\widetilde{\bfx}$ is regular. Therefore, the two deletions in $\bfx$ is within a window of length at most $7\log n+1$.

    For $\sigma\in\Sigma$, by calculating $N_\sigma(\bfx)-N_\sigma(\bfx')\bmod{3}$, the values of two deleted symbols can be obtained. If there is a $\sigma\in\Sigma$, such $N_\sigma(\bfx)-N_\sigma(\bfx')\bmod{3}=t$, then the number of $\sigma$ deleted from $\bfx$ is $t$. We consider the two cases whether the two deleted symbols are the same.
    \begin{itemize}
        \item The two deleted symbols are $\sigma_1\neq \sigma_2$:\\
        We prove that inserting $\sigma_1$ and $\sigma_2$ into $\bfx'$ can only get $\bfx$ if we want to keep $\Sum(\bfx^\sigma)=b_\sigma\bmod{14\log n}$ for all $\sigma\in\Sigma$. Otherwise, suppose there is another way to insert $\sigma_1$ and $\sigma_2$ satisfying the constraints, and the resultant sequence is denoted as $\bfy$. Let $i_1$ and $i_2$ denote the indices of the deleted $\sigma_1$ and $\sigma_2$ in $\bfx$, and $\{j_1,j_2\}$ denote the set of indices of the inserted $\sigma_1$ and $\sigma_2$ in $\bfy$. Without loss of generality, we assume $i_1\leq j_1$, $i_1< i_2$, and $j_1 <j_2$. By Lemma \ref{lem:del01}, we have $i_2=i_1+\epsilon_1$ and $j_2=j_1+\epsilon_2$, where $\epsilon_1,\epsilon_2\in\{1,2\}$. We claim that $j_2>i_2$. Otherwise if $j_2\leq i_2$, then to make sure $\bfx\neq \bfy$ we must have $\epsilon_1=2$, and $j_2=i_2=j_1+1$ or $j_1=i_1=j_2-1$. In the case of $j_2=i_2=j_1+1$, we have $\bfx_{[i_1,i_2]}=\sigma_1 x_{i+1}\sigma_2$, and $\bfy_{[i_1,i_2]}=x_{i+1}\sigma_1\sigma_2$ or $x_{i+1}\sigma_2\sigma_1$. If $x_{i+1}\leq \sigma_2$, then $\sigma_1>x_{i+1}$ by the condition that $\widetilde{\bfx}_{[i_1,i_2]}$ is alternate. So $\bfy_{[i_1,i_2]}$ cannot be $x_{i+1}\sigma_1\sigma_2$ or $x_{i+1}\sigma_2\sigma_1$, where both are contradictory to $\widetilde{\bfx}_{[i_1,i_2]}=\widetilde{\bfy}_{[i_1,i_2]}$. If $x_{i+1}>\sigma_2$, then $ \sigma_1\leq x_{i+1}$, and due to $\widetilde{\bfy}_{i_1}=\widetilde{\bfx}_{i_1}=1$ the only possible of $\bfy_{[i_1,i_2]}$ will be $x_{i+1}\sigma_1\sigma_2$, where $x_{i+1}=\sigma_1$. This leads to $\bfx=\bfy$, again a contradiction. The case $j_1=i_1=j_2-1$ can be proved similarly, so we have $j_2>i_2$.
        If $\bfx_{[i_1,j_2]}$ consists of $\sigma_1$ and $\sigma_2$ only, then to keep $\widetilde{\bfx}_{[i_1,j_2-1]}$ is an alternate sequence it has to be $\bfx=\bfy$ by Lemma \ref{lem:unique quaternary}. Thus, there must exist a $\sigma_3\in\Sigma\setminus\{\sigma_1,\sigma_2\}$ in $\bfx_{[i_1,j_2]}$. Due to $\bfx_{[i_1,j_2]\setminus\{i_1,i_2\}}=\bfy_{[i_1,j_2]\setminus\{j_1,j_2\}}$ and $i_1<i_2<j_2$, we have
        $$N_{\sigma_3}(\bfx_{[i_1,j_2]})\leq\Sum(\bfx^{\sigma_3})-\Sum(\bfy^{\sigma_3})\leq 2N_{\sigma_3}(\bfx_{[i_1,j_2]}),$$ which is a contradiction to $\Sum(\bfx^{\sigma_3})-\Sum(\bfy^{\sigma_3})=0\bmod{14\log n}$.
        \item The two deleted symbols are both equal to $\sigma_1$:\\
        Like the previous case, we prove that inserting these two $\sigma_1$ into $\bfx'$ can only get $\bfx$ if we want to keep $\Sum(\bfx^\sigma)=b_\sigma\bmod{14\log n}$ for all $\sigma\in\Sigma$. By Lemma \ref{lem:del01}, the two $\sigma_1$ inserted in $\bfx'$ are consecutive or separated by one position. Suppose there is another way to insert two $\sigma_1$ into $\bfx'$, and the resultant sequence is $\bfy\in\cC_2^D$. We again assume the indices of these two $\delta_1$ in $\bfx$ and $\bfy$ are $i_1<i_2$ and $j_1<j_2$, respectively. Without loss of generality, let $i_1<j_1$. Since $\bfx\neq\bfy$, there must be a $\sigma_2\in\Sigma\setminus\{\delta_1\}$ in $\bfx_{[i_1,j_2]}$. We can calculate that
        $$N_{\sigma_2}(\bfx_{[i_1,j_2]})\leq\Sum(\bfx^{\sigma_2})-\Sum(\bfy^{\sigma_2})\leq 2N_{\sigma_2}(\bfx_{[i_1,j_2]}),$$
        which contradicts that $\Sum(\bfx^{\sigma_2})\equiv\Sum(\bfy^{\sigma_2})\bmod{14\log n}$.
    \end{itemize}
    Now we have proved that we can uniquely recover $\bfx$ from two deletions if the two deletions in $\widetilde{\bfx}$ are consecutive $01$ or $10$ in an alternate substring. Combining with Lemma \ref{lem:liu2024}, the code $\cC_2^D$ is indeed a quaternary two-deletion correcting code.

    For the redundancy, there exist parameters, such that the size of $\cC_2^D$ is at least
    $$\frac{4^n}{c\cdot n^4\cdot (\log n)^{10}\cdot 4n\cdot 3^4\cdot (14\log n)^4},$$
    where $c$ is a constant and from $\cC_2$. Therefore, the redundancy is at most $5\log n+14\log\log n+O(1)$.
\end{IEEEproof}

\begin{construction}\label{con:2 syndef code}
    For $\bfc\in\ZZ_{14\log n}^{(M-4M_1)\times 4}$, $\bfP=(P_1,P_2)$, where $P_1,P_2\leq 28\log n+5$, we denote
    \begin{align*}
        \cC_2^{SD}=& \Big\{\sC\in(\Sigma^n)^M: \bfc_i=S_{a_i}(\bfx_i) \text{ for }i\in[4M_1] \text{ where }\bfx_i\in\cC_2^D,\\
        & \widetilde{\bfc_i}\in\cC_2^\bfP, \text{ and } \Sum(\bfc_i^j)=b_{i,j}\bmod{14\log n}\\
        & \text{for } i\in[4M_1+1,M], j\in\Sigma\Big\},
    \end{align*}
    where for $1\leq i\leq 4M_1$, $a_i$ is chosen based on Corollary \ref{cor:cover}.
\end{construction}

\begin{lemma}
    $\cC_2^\bfP$ is not only a binary $\bfP$-bounded two-deletion correcting code, but also a single-deletion correcting code.
\end{lemma}
\begin{IEEEproof}
    From the construction of $\cC_2^\bfP$, every codeword in $\cC_2^\bfP$ has the form $(\bfz,0,1,\cE_1(\bfz),\cE_2(\bfz),\xi(\cE_1(\bfz),\cE_2(\bfz)))$, where $\bfz\in\{0,1\}^k$ and $\cE_1,\cE_2,\xi$ are labeling function we will describe in Appendix \ref{sec:large P bounded}. As we will show, $\xi(\bfz)$ allows us recover $\bfz$ from two deletions from the erroneous sequence, and $\cE_2$ contains the value $\VT(\bfz)\bmod k+1$.

    When a deletion occurs in a sequence with the form $(\bfz,0,1,\cE_1(\bfz),\cE_2(\bfz),\xi(\cE_1(\bfz),\cE_2(\bfz)))$ and result in $\bfx\in\{0,1\}^{n-1}$, we first look at the $k+1$ position of $\bfx$. If $x_{k+1}$=0, then there is no error in $\bfz$, so we can calculate the values $\cE_1(\bfx_{[k]}),\cE_2(\bfx_{[k]}),\xi(\cE_1(\bfx_{[k]}),\cE_2(\bfx_{[k]}))$, and return $(\bfx_{[k]},0,1,\cE_1(\bfx_{[k]}),\cE_2(\bfx_{[k]}),\xi(\cE_1(\bfx_{[k]}),\cE_2(\bfx_{[k]})))$ as the correct sequence. If $x_{k+1}=1$, then the deletion occurs at $\bfz$. We use $\bfx_{[k-1]}$ and the value $\VT(\bfz)\bmod k+1$ in $\cE_2(\bfz)$ to recover $\bfz$, then we return $(\bfz,\bfx_{[k,n]})$ as the correct sequence.
\end{IEEEproof}

\begin{theorem}\label{thm:2 syndef code}
    $\cC_2^{SD}$ is a $2$-SDCC, and there exist parameters such that the redundancy is at most $\frac{12}{2-\log 3}(\log n)^2+M(2\log n+26\log\log n+o(\log\log n))-\Omega(\log n\log\log n)$.
\end{theorem}

\begin{IEEEproof}
    The proof is similar to that of Theorem~\ref{thm:1sdcc}.
\end{IEEEproof}

\section{A Lower Bound on the Redundancy of Known-Synthesis-Defect Correcting Codes}\label{sec:Bound}
In this section, we give the lower bound for redundancy on $1$-KDCC using some tools from graph theory. Specifically, we use the same method as \cite[Section~IV]{chrisnata2022correcting} to derive the lower bound for redundancy.
First, let us introduce some definitions.

\begin{definition}\label{def:cliquecover}
  A collection $\cQ$ of cliques is a clique cover of a graph $\cG$ if every vertex in $\cG$ belongs to some clique in $\cQ$.
\end{definition}

\begin{lemma}[\hspace*{-1.2mm}\cite{knuth1994sandwich}]\label{lem:clique}
  If $\cQ$ is a clique cover of $\cG$, then the size of any independent set of $\cG$ is at most $|\cQ|$.
\end{lemma}

Let the vertices of $\cG$ to be all quaternary sequences of length $n$, and two vertices $\bfu,\bfv$ are connected if there exists a set $\Delta=\{\delta\}$, such that $\syndef_{\Delta}(\bfu)=\syndef_{\Delta}(\bfv)$, where $\delta\in[4n]$. Obviously, an independent set of $\cG$ is a $1$-KDCC. By Lemma \ref{lem:clique}, the upper bound of the size of $1$-KDCC is upper bounded by the size of any one of clique cover of $\cG$. Now we give a construction of a clique cover of $\cG$.

\begin{lemma}\label{lem:size of cliquecover}
  There exists a clique cover $\cQ$ for $\cG$ and is of size $4^{n-1}(1+3(\frac{4^5-16}{4^5})^{\lfloor\frac{n}{5}\rfloor})$.
\end{lemma}
\begin{IEEEproof}
    Let
\begin{align*}
   & A=\{11234,12134,12314,12341\} \\
   & B=\{22341,23241,23421,23412\} \\
   & C=\{33412,34312,34132,34123\} \\
   & D=\{44123,41423,41243,41234\},
\end{align*}
and $E=\Sigma^5\setminus(A \cup B \cup C \cup D)$.
Furthermore, define $$Z=\left\{(\bfp,\bfr,i):\bfp\in E^i,\bfr\in\Sigma^{n-5i-5},i\in\{0,\ldots,\lfloor\frac{n}{5}\rfloor-1\}\right\}.$$
For every $\bfz=(\bfp,\bfr,i)\in Z$, we define
\begin{align*}
   & Q_{\bfz}^{(A)}=\{\bfp\bfq\bfr:\bfq\in A\}, \\
   & Q_{\bfz}^{(B)}=\{\bfp\bfq\bfr:\bfq\in B\}, \\
   & Q_{\bfz}^{(C)}=\{\bfp\bfq\bfr:\bfq\in C\}, \\
   & Q_{\bfz}^{(D)}=\{\bfp\bfq\bfr:\bfq\in D\}
\end{align*}
to be the cliques of size 4. For the remaining vertices, we define
$$S_\bfx=\{\bfx\}, \text{ where } \bfx\in E^{\lfloor\frac{n}{5}\rfloor}\times\Sigma^{n-5\lfloor\frac{n}{5}\rfloor}$$
to be the singletons.

Define
\begin{align*}
    \cQ= & \left\{Q_{\bfz}^{(A)},Q_{\bfz}^{(B)},Q_{\bfz}^{(C)},Q_{\bfz}^{(D)},\bfz\in Z\right\}\cup\\
         & \left\{Q_\bfx:\bfx\in E^{\lfloor\frac{n}{5}\rfloor}\times\Sigma^{n-5\lfloor\frac{n}{5}\rfloor}\right\}.
\end{align*}

    To prove that $\cQ$ is a clique cover, we need to show every set in $\cQ$ is a clique and all vertices in these sets cover the whole space $\Sigma^n$.

    First, for every $\bfx\in E^{\lfloor\frac{n}{5}\rfloor}\times\Sigma^{n-5\lfloor\frac{n}{5}\rfloor}$, the set $Q_\bfx$ has only one element and thus is a clique.

    Then, for $\bfz\in Z$, the set $Q_{\bfz}^{(A)}$ has $4$ elements, which are $\bfp11234\bfr,\bfp12134\bfr,\bfp12314\bfr,\bfp12341\bfr$. Suppose $\cycle(\bfp11234\bfr)_{|\bfp|+2}=\delta$, then $\syndef_{\{\delta\}}(\bfp11234\bfr)=\syndef_{\{\delta\}}(\bfp12134\bfr)=\syndef_{\{\delta\}}(\bfp12314\bfr)=\syndef_{\{\delta\}}(\bfp12341\bfr)$, so they are connected to each other and thus $Q_{\bfz}^{(A)}$ is a clique. Similarly, $Q_{\bfz}^{(B)}$,$Q_{\bfz}^{(C)}$,$Q_{\bfz}^{(D)}$ are all cliques.

    For a vertex $\bfv\in\Sigma^n$, if there exists a $i\in\{0,\ldots,\lfloor\frac{n}{5}\rfloor-1\}$, such that $v_{[5i+1,5i+5]}\in A\cup B \cup C \cup D$, then $\bfv\in Q_\bfz^{(A)}$ or $Q_\bfz^{(B)}$ or $Q_\bfz^{(C)}$ or $Q_\bfz^{(D)}$ for some $\bfz=(\bfp,\bfr,i)$. If no such $i$ exists, then $\bfv\in S_\bfv$. Therefore, we conclude that $\cQ$ is a clique cover.

    For the size of $\cQ$, we calculate as follows. For every $\bfz\in Z$, there are $4$ cliques of $Q_{\bfz}^{(A)},Q_{\bfz}^{(B)},Q_{\bfz}^{(C)}$ and $Q_{\bfz}^{(D)}$. The size of $Z$ is
    $$\sum_{i=0}^{\lfloor\frac{n}{5}\rfloor-1}(4^5-16)^i4^{n-5i-5}.$$
    So there are total $4|Z|$ cliques of size $4$. On the other side, the size of singleton is $$(4^5-16)^{\lfloor\frac{n}{5}\rfloor}4^{n-5\lfloor\frac{n}{5}\rfloor}.$$
    Therefore,
    \begin{align*}
        |\cQ|&=4\sum_{i=0}^{\lfloor\frac{n}{5}\rfloor-1}(4^5-16)^i4^{n-5i-5}+(4^5-16)^{\lfloor\frac{n}{5}\rfloor}4^{n-5\lfloor\frac{n}{5}\rfloor}\\
             &=4^{n-1}(1+3(\frac{4^5-16}{4^5})^{\lfloor\frac{n}{5}\rfloor}).
    \end{align*}
\end{IEEEproof}

\begin{theorem}
    Let $\cC\subseteq\Sigma^n$ be a $1$-KDCC, then the redundancy of $\cC$ is at least $\log4-o(1)$.
\end{theorem}
\begin{IEEEproof}
    An independent set of $\cG$ has the property that for any $\bfu,\bfv\in\cG$, $\syndef_{\{\delta\}}(\bfu)\cap\syndef_{\{\delta\}}(\bfv)=\varnothing$ for $\delta\in[4n]$. This means an independent set of $\cG$ is a $1$-KDCC. By Lemma \ref{lem:clique}, the size of any independent set is upper bounded by the size of a clique cover of $\cG$, and we can construct a clique cover of size $4^{n-1}(1+3(\frac{4^5-16}{4^5})^{\lfloor\frac{n}{5}\rfloor})$ by Lemma \ref{lem:size of cliquecover}. So the size of a $1$-KDCC is upper bounded by $4^{n-1}(1+3(\frac{4^5-16}{4^5})^{\lfloor\frac{n}{5}\rfloor})$, and the redundancy of it is at least $n-\log(4^{n-1}(1+3(\frac{4^5-16}{4^5})^{\lfloor\frac{n}{5}\rfloor}))=1-o(1)$.
\end{IEEEproof}

With this theorem, it follows that the $1$-KDCC $\cC_1^{KDCC}(n;a)$ we construct is almost optimal.

\section{Conclusion and Future Work}\label{sec:future}
We investigate codes correcting synthesis defects and provide two type of codes to correct this error. In the first one, we assume we have already known the information of defective cycles. Although this seems to provide sufficient information about the errors, it is not enough to recover the strand. We construct $t$-KDCCs for $t=1,2$, and this shows the strategy to construct $t$-KDCC for general $t$. Besides, we derive a bound on the size of $1$-KDCC, which shows our construction for $1$-KDCC is almost optimal. In the second type of codes, each codeword is a set of $M$ sequence, and they share the same synthesis defects. We utilize this feature, and provide $t$-SDCCs for $t=1,2$, where each codeword is divided into two parts with different coding schemes. The redundancy of our $t$-SDCCs is roughly $\lambda_1(\log n)^2+M\log\log n$ and $\lambda_2(\log n)^2+2M\log n$ for $t=1,2$, repectively, for some constants $\lambda_1$ and $\lambda_2$. In contrast, if we employ the best existing single deletions and two-deletion correcting codes for each sequence, then the redundancy for $1$ and $2$-SDCC will be at least $M\log n$ and $4M\log n$.

As explained in Section \ref{subsec:reduct to binary}, if we want to construct a $t$-KDCC, then if suffices to construct a binary $\bfP$-bounded $t$-deletion correcting code. We only provide this code for $t=2$ in this paper, and the cases for $t>2$ are left to the future.

For the $t$-SDCC, we only provide the constructions for $t\leq2$. The case $t\geq 3$ is far more complicated. Even if we locate each defect in a small interval, we cannot tell which cycles are defective in a sequence. For example, three cycles $\delta_1,\delta_2,\delta_3$ are defective, but if there is a sequence suffers from $2$ deletions, then we do not know these two deletions are caused by $\delta_1,\delta_2$ or $\delta_2,\delta_3$ or $\delta_1,\delta_3$. So, it is not enough to let it belongs to only a $(P_1,P_2,P_3)$-bounded $3$-deletion correcting code. Even if it also belongs to a $(P_1,P_2)$-bounded $2$-deletion correcting code, we cannot recover it because we do not know the two interval. The problem of constructing $t$-SDCCs for $t\geq 3$ is left for future work.

\bibliographystyle{IEEEtran}
\bibliography{references}

\appendix
\section{Construction of P-bounded two-deletion correcting code for large P}\label{sec:large P bounded}

To correct two deletions where each deletion is within an interval of length $P_i$, we distinguish if the two intervals are adjacent or not.

~\\
\noindent 1) When the two intervals are adjacent or intersect

Denote $\rho=P_1+P_2$ and
\begin{equation*}
  I_i=\begin{cases}
        [(i-1)\rho+1,(i+1)\rho], & \mbox{for } i\in\{1,\ldots,\lceil n/\rho\rceil-2\}, \\
        [(i-1)\rho+1,n], & \mbox{for } i=\lceil n/\rho\rceil-1.
      \end{cases}
\end{equation*}

\begin{lemma}[\hspace*{-1.2mm}{\cite[Theorem 1]{sima2019}}]\label{lem:2del 7logn}
  For any integer $n\geq 3$, there exist a sketch function $\xi:\{0,1\}^n\rightarrow\{0,1\}^{r_0(n)}$, computable in linear time, such that for any $\bfx\in\{0,1\}^n$, given $\xi(\bfx)$ and $\bfx_{[n]\setminus\{i,j\}}$, one can uniquely recover $\bfx$, where $i,j\in[n]$, where $r_0(n)=7\log n+O(\log\log n)$.
\end{lemma}

\begin{construction}
Define $\cE_1:\{0,1\}^n\rightarrow\{0,1\}^{r_1(P_1,P_2)}$ as follows:\\
for $\bfx\in\{0,1\}^n$,
\begin{align*}
    \cE_1(\bfx)=&\Big(\sum_{j=1}^{\lceil\frac{\lceil n/\rho\rceil-1}{2}\rceil}\xi(\bfx_{I_{2j-1}}) \bmod 2^{7\log\rho+O(\log\log\rho)},\\
    &\sum_{j=1}^{\lfloor\frac{\lceil n/\rho\rceil-1}{2}\rfloor}\xi(\bfx_{I_{2j}}) \bmod 2^{7\log\rho+O(\log\log\rho)}\Big),
\end{align*}
where $r_1(P_1,P_2)=14\log\rho+O(\log\log\rho)$
\end{construction}

\begin{lemma}\label{lem:adjacent interval}
  For any integers $P_1,P_2\geq2$ and $n\geq3$, there exists a sketch function $\cE_1:\{0,1\}^n\rightarrow\{0,1\}^{r_1(P_1,P_2)}$, computable in linear time, such that for any $\bfx\in\{0,1\}^n$, if $\bfx'$ is obtained by deleting two bits at two adjacent or overlapping intervals of lengths $P_1,P_2$, then given $\cE_1(\bfx)$ and $\bfx'$ together with the locations of this two intervals, we can uniquely recover $\bfx$, where $r_1(P_1,P_2)=14\log(P_1+P_2)+O(\log\log(P_1+P_2))$
\end{lemma}
\begin{IEEEproof}
    As $\bfx'$ is obtained by deleting two bits at two adjacent or overlapping intervals of lengths $P_1,P_2$, the locations of two deletions are within an interval $I$ of length $\rho=P_1+P_2$. Furthermore, there exists a $j\in[\lceil\frac{n}{\rho}\rceil-1]$, such that $I\subseteq I_j$. Since $I_j=[(j-1)\rho+1,(j+1)\rho]$, we know $\bfx'_{[(j-1)\rho+1,(j+1)\rho-2]}$ is a subsequence of $\bfx_{[(j-1)\rho+1,(j+1)\rho]}$, and $\bfx_{[n]\setminus[(j-1)\rho+1,(j+1)\rho]}=\bfx'_{[n-2]\setminus[(j-1)\rho+1,(j+1)\rho-2]}$. So we can get $\xi(\bfx_{I_j})\bmod 2^{7\log\rho+O(\log\log\rho)}$ from $\cE_1(\bfx)$. By Lemma \ref{lem:2del 7logn}, given $\xi(\bfx_{I_j})$, $\bfx'_{[(j-1)\rho+1,(j+1)\rho-2]}$, we can uniquely recover $\bfx_{I_j}$, so $\bfx$ is recovered.
\end{IEEEproof}

~\\
\noindent 2) When the two intervals are separated by at least one symbol

We use the same analytical approach as in \cite{guruswami2021}.
First, we use $f_0(\bfx)=\sum_{i=1}^{n}x_i\bmod3$ to identify the following 3 cases:
\begin{itemize}
  \item[i)] the two deleted symbols are both $0$;
  \item[ii)] the two deleted symbols are both $1$;
  \item[iii)] the two deleted symbols are $1$ and $0$.
\end{itemize}

Besides, we need two more functions:
$$f_1(\bfx)=\sum_{i=1}^{n}ix_i,$$
and $$f_2(\bfx)=\sum_{i=1}^{n}\binom{i}{2}x_i.$$

\begin{lemma}[\hspace*{-1.2mm}{\cite[Lemma 5]{guruswami2021}}]\label{lem:00or11}
If we have deleted two $1$’s or two $0$’s in forming
the subsequence $\bfx'$ from $\bfx$, then $f_1(\bfx)$ and $f_2(\bfx)$ together with $\bfx'$ identify $\bfx$ uniquely.
\end{lemma}

\begin{lemma}[\hspace*{-1.2mm}{\cite[Observation 2]{guruswami2021}}]\label{lem:01}
If we have deleted a $0$ and a $1$ from two nonadjacent intervals respectively in forming
the subsequence $\bfx'$ from $\bfx$, then $f_1(\bfx)$ and $f_2(\bfx)$ together with $\bfx'$ identify $\bfx$ uniquely.
\end{lemma}
\begin{IEEEproof}
As in \cite{guruswami2021}, we insert a $0$ and a $1$ as far right as possible in the two intervals of $\bfx'$ to make the value of $f_1(\bfx)$ correct.

When we move the $0$ left passing a $1$, the value of $f_1(\bfx)$ increases $1$, so we need to move the $1$ in the other interval left passing a $0$ to keep the value of $f_1(\bfx)$ constant. During this process, the order of the $0$ and $1$ will keep, since the moving bits in the left interval will never move to the right interval. So by Lemma 2 and Observation 2 of \cite{guruswami2021}, there will never happen an overtake event, so $f_2(\bfx)$ will be monotonically increasing or decreasing in the process.
\end{IEEEproof}

Combining Lemma \ref{lem:00or11} and \ref{lem:01}, we can prove that if we have $\bfx'$ together with $f_0(\bfx),f_1(\bfx),f_2(\bfx)$, then $\bfx$ can be uniquely recover knowing the two nonadjacent intervals of two deletions located. Now we need to determine how many values do we need for $f_1(\bfx)$ and $f_2(\bfx)$.

Since the two deleted bits can only move within their intervals, the bits outside these two intervals will keep their contributions to $f_1(\bfx)$ and $f_2(\bfx)$. So it suffices to consider the bits in these two intervals.

First we consider $f_1(\bfx)$. We analysize case by case:
\begin{itemize}
  \item[i)] the two deleted symbols are both $0$: For the $i$th interval, its contribution to $f_1(\bfx)$ will increase at most $P_i-1$ during the process of $0$ moving left. So $f_1(\bfx)$ will increase at most $P_1+P_2-2$.
  \item[ii)] the two deleted symbols are both $1$: For the $i$th interval, its contribution to $f_1(\bfx)$ will increase at most $P_i-1$ during the process of $1$ moving rihgt. So $f_1(\bfx)$ will increase at most $P_1+P_2-2$.
  \item[iii)] the two deleted symbols are $1$ and $0$: The value of $f_1(\bfx)$ is least when $1$ is inserted in the front of the first interval and $0$ is inserted in the end of the second interval, and is largest when $0$ is inserted in the front of the first interval and $1$ is inserted in the end of the second interval. The difference of them is $j_2+P_2-1-j_1<n$, where $j_i$ is the first index of $i$th interval.
\end{itemize}
Therefore, it has at most $n$ values for $f_1(\bfx)$, so it suffices to apply modular $n$.

Now we consider $f_2(\bfx)$. Since we must keep $f_1(\bfx)$ constant when we insert the two bits in $\bfx'$, we do not have to consider all the possibilities of insertions. There are at most $\min\{P_1,P_2\}$ possibilities of insertions to keep $f_1(\bfx)$ constant. By \cite[Lemma 2]{guruswami2021}, we can easily identify the extremal values of $f_2(\bfx)$.
\begin{itemize}
  \item[i)] the two deleted symbols are both $0$: The value of $f_2(\bfx)$ will increase by $\binom{j+1}{2}-\binom{j}{2}=j$ if the moving $0$ moves left passing a $1$ located at $j$, and will decrease by $\binom{i}{2}-\binom{i-1}{2}=i-1$ if the moving $0$ moves right passing a $1$ located at $i$. First we insert the two $0$'s as far away from each other as possible to keep $f_1(\bfx)$ correct. Then, every time we move the left $0$ right passing a $1$ located at $i$, the right $0$ need to be moved left passing a $1$ located at $j$. In this process, $f_2(\bfx)$ will increase $j-i+1$. Suppose the indices of $1$'s in the left interval are $\{i_1,\ldots,i_p\}$, and in the right interval are $\{j_1,\ldots,j_q\}$, where $p<P_1$ and $q<P_2$. When the two moving $0$'s are as far close to each other as possible, the value of $f_2(\bfx)$ grows to maximum, and will increase at most $\sum_{k=1}^{\min\{p,q\}}(j_k-i_k+1)<\min\{P_1,P_2\}n$ compared to the beginning.
  \item[ii)] the two deleted symbols are both $1$: The value of $f_2(\bfx)$ will decrease by $\binom{i+1}{2}-\binom{i}{2}=i$ if the moving $1$ moves left passing a $0$ located at $i$, and will increase by $\binom{j}{2}-\binom{j-1}{2}=j-1$ if the moving $1$ moves right passing a $0$ located at $j$. First we insert the two $1$'s as far close to each other as possible to keep $f_1(\bfx)$ correct. Then, every time we move the left $1$ left passing a $0$ located at $i$, the right $1$ need to be moved right passing a $0$ located at $j$. In this process, $f_2(\bfx)$ will increase $j-i-1$. Suppose the indices of $0$'s in the left interval are $\{i_1,\ldots,i_p\}$, and in the right interval are $\{j_1,\ldots,j_q\}$, where $p<P_1$ and $q<P_2$. When the two moving $1$'s are as far away from each other as possible, the value of $f_2(\bfx)$ grows to maximum, and will increase at most $\sum_{k=1}^{\min\{p,q\}}(j_k-i_k-1)<\min\{P_1,P_2\}n$ compared to the beginning.
  \item[iii)] the two deleted symbols are $1$ and $0$: By Lemma \ref{lem:01}, The moving bits can never change their intervals. So we can consider the two case separately.
      \begin{itemize}
        \item $0$ is inserted at the first interval, and $1$ the second. Similar to the above two cases, we insert $0$ and $1$ as left as possible to keep $f_1(\bfx)$ correct. Once we move $0$ right passing a $1$ located at $i$, $1$ need to be moved right passing a $0$ located at $j$. $f_2(\bfx)$ will increase by $j-i$ in this process. When the two moving
        $0$ and $1$ are as far right as possible, the value of $f_2(\bfx)$ will increase at most $\min\{P_1,P_2\}n-1$.
        \item $1$ is inserted at the first interval, and $0$ the second. Similar to the previous, $f_2(\bfx)$ will increase at most $\min\{P_1,P_2\}n-1$.
      \end{itemize}
\end{itemize}
Therefore, it has at most $\min\{P_1,P_2\}n$ values for $f_2(\bfx)$, so it suffices to apply modular $\min\{P_1,P_2\}n$ for $f_2(\bfx)$.

\begin{construction}
Define $\cE_2:\{0,1\}^n\rightarrow\{0,1\}^{r_2(n,P_1,P_2)}$ as follows:\\
for $\bfx\in\{0,1\}^n$,
\begin{align*}
    \cE_2(\bfx)=\Big(& f_0(\bfx)\bmod 3,f_1(\bfx)\bmod n+1,\\
     & f_2(\bfx)\bmod Pn\Big),
\end{align*}
where $r_2(n,P_1,P_2)=2\log n+\log P+O(1)$.
\end{construction}

By Lemma \ref{lem:00or11} and \ref{lem:01}, and the above analysis for ranges of $f_1(\bfx)$ and $f_2(\bfx)$, we have the following lemma.

\begin{lemma}\label{lem:separated interval}
    For any integers $P_1,P_2\geq2$ and $n\geq3$, there exists a sketch function $\cE_2:\{0,1\}^n\rightarrow\{0,1\}^{r_2(n,P_1,P_2)}$, computable in linear time, such that for any $\bfx\in\{0,1\}^n$, if $\bfx'$ is obtained by deleting two bits at two non-overlapping intervals of lengths $P_1,P_2$, then given $\cE_2(\bfx)$ and $\bfx'$ together with the locations of this two intervals, we can uniquely recover $\bfx$, where $r_2(n,P_1,P_2)=2\log n+\log P+O(1)$.
\end{lemma}

\begin{construction}\label{con:E}
  Define $\cE:\{0,1\}^n\rightarrow\{0,1\}^{n+r(n,P_1,P_2)}$ as follows:\\
for $\bfx\in\{0,1\}^n$,
$$\cE(\bfx)=\Big(\bfx,\cE_1(\bfx),\cE_2(\bfx),\xi(\cE_1(\bfx),\cE_2(\bfx))\Big),$$ where $r(n,P_1,P_2)=r_1(P_1,P_2)+r_2(n,P_1,P_2)+r_0(r_1(P_1,P_2)+r_2(n,P_1,P_2))=2\log n+7\log\log n+14\log(P_1+P_2)+\log P+O(\log\log(P_1+P_2))$.
\end{construction}

\begin{IEEEproof}[Proof of Lemma \ref{lem:bounded two del}]
    To be convenient, we omit the parameters of $r_0,r_1,r_2$ and $r$, and use them to denote the lengths of $\xi(\cE_1(\bfx),\cE_2(\bfx)),\cE_1(\bfx),\cE_2(\bfx)$ and $\cE(\bfx)$, respectively.
    Since we know the information of the two intervals, we can distinguish if the two intervals are separated.
    \begin{itemize}
        \item If the intervals do not intersect with $[n]$: we know the part of $\bfx$ is correct, so just set $\bfx=\cE'(\bfx)_{[n]}$.
        \item If the intervals intersect with $[n]$:
        \begin{itemize}
            \item if the two intervals are not separated, we first recover $\cE_1(\bfx),\cE_2(\bfx)$ from $\cE(\bfx)_{[n+r_1+r_2+1,n+r-2]}$ and $\cE(\bfx)_{[n+1,n+r_1+r_2-2]}$. It is obvious that $\cE(\bfx)_{[n+r_1+r_2+1,n+r-2]}$ is a subsequence of $\xi(\cE_1(\bfx),\cE_2(\bfx))$ of length $r_0-2$, and $\cE(\bfx)_{[n+1,n+r_1+r_2-2]}$ is a subsequence of $\cE_1(\bfx),\cE_2(\bfx)$ of length $r_1+r_2-2$. By Lemma \ref{lem:2del 7logn}, we can recover $\cE_1(\bfx),\cE_2(\bfx)$. Then, we can recovery $\bfx$ by using $\cE(\bfx)_{[1,n-2]}$ and $\cE_1(\bfx)$ by Theorem \ref{lem:adjacent interval}.
            \item if the two intervals are separated, similar to above, we can recover $\cE_1(\bfx),\cE_2(\bfx)$, and we use $\cE(\bfx)_{[1,n-2]}$ and $\cE_2(\bfx)$ to recover $\bfx$ by Theorem \ref{lem:separated interval}.
        \end{itemize}
    \end{itemize}
\end{IEEEproof}

\end{document}